\documentclass[%
 aapm,
 mph,%
 amsmath,amssymb,
 reprint,%
]{revtex4-2}

\usepackage{graphicx}
\usepackage{dcolumn}
\usepackage{bm}
\usepackage{graphicx}
\usepackage{subcaption}
\usepackage{afterpage}
\usepackage{placeins}
\usepackage{ragged2e}
\usepackage[font=small,labelfont=bf,
justification=justified,
format=plain]{caption}
\begin{document}

\preprint{AAPM/123-QED}

\title{Dynamics of Multi-Domains in Ferroelectric Tunnel Junction}

\author{Nilesh Pandey}
 \author{Yogesh Singh Chauhan}
 \email{chauhan@iitk.ac.in}
  \email{pandeyn@iitk.ac.in}
\affiliation{ 
Department of Electrical Engineering, Indian Institute of Technology - Kanpur,  Kanpur, 208016, India
}

\begin{abstract}
\textbf{The discovery of giant tunnel electroresistance (TER) in Ferroelectric Tunnel Junction (FTJ) paves a futuristic possibility of utilizing the FTJ as a bi-stable resistive device with an enormously high ON/OFF ratio. In the last 20 years numerous studies have reported that the formation of multi-domain in ferroelectric material is an inevitable process to minimize the total system energy. Recent studies based on phase field simulations have demonstrated that domain nucleation/motion substantially alters the electrostatics of a ferroelectric material. However, the impact of domain dynamics on quantum transport in FTJ remains elusive. This paper presents a comprehensive study of multi-domain dynamics in a ferroelectric tunnel junction. The analysis of this article is twofold; firstly, we study the impact of domain dynamics on electrostatics in an FTJ. Subsequently, the obtained electrostatics is used to study the variations in tunneling current, and TER originated from multi-domain dynamics. We show that ON/OFF current density and TER vary locally in the ferroelectric region. Furthermore, due to the periodic domain texture, device's electrostatics and quantum transport exhibit an oscillatory nature. The ON/OFF current density shows a sine/cosine distribution in ferroelectric, and approximately one-decade local variations in current density are observed. These local fluctuations in current density cause the oscillations in the device's ON/OFF ratio (TER). The optimization techniques to achieve a uniform and maximum TER are also discussed.
On top of that, a 2-D analytical and explicit model is derived by solving coupled 2-D Poisson's equation and Landau-Ginzburg equation. The switching and nucleation of domains are incorporated in the model by minimizing net ferroelectric energy (depolarization energy density+free energy density+gradient energy density). Furthermore, the impact of the bottom insulator layer on ferroelectric's gradient energy is also studied. }

\end{abstract}

\maketitle

\section*{Introduction}
A ferroelectric tunnel junction (FTJ) utilizes the mechanism of quantum tunneling to switch between bi-stable states \cite{Velev J. P}$^-$\cite{Dong Z Cao}. To explore the microscopic physics of quantum tunneling in FTJ, first principle-based atomistic/numerical simulations were performed \cite{Velev J. P}$^,$\cite{Caffrey}$^,$\cite{Bilc}$^,$\cite{Borisov V S}$^,$\cite{Quindeau A et}.
The ON/OFF characteristic of FTJ is defined by modulation in electrical conductance due to polarization switching. Subsequently, by utilizing the ON/OFF ratio of current density, tunnel electroresistance (TER) is calculated\cite{Velev J. P}$^,$\cite{Garcia V_1}$^,$\cite{Gruverman A et al}$^,$\cite{Bilc}$^,$\cite{Garcia_3}$^,$\cite{Xi Z}. Moreover, many types of research have demonstrated that the ability of room temperature restive switching in FTJ exhibits a profound candidacy of the FTJ to utilize as a non-volatile memory component\cite{Gruverman A et al}$^,$\cite{Pantel D Goetze.}$^,$\cite{Chanthbouala A}$^,$\cite{Radaelli G. et}.
Furthermore, the researchers have studied epitaxial fabricated FTJs to achieve a higher value of the TER\cite{Zenkevich A e}$^,$\cite{Li Z}$^,$\cite{Soni R et a}$^,$\cite{Sokolov A Bak}. On the other hand, various researches demonstrated that the TER of FTJ can be tuned by altering the potential barrier or by engineering the electrical junctions\cite{Lu H}$^,$\cite{BoynSet}$^,$\cite{Li C et al}$^,$\cite{Liu X Burton}$^,$\cite{Xi Z}.
Recently, the gigantic value of TER $\approx$ 10$^7$ is experimentally achieved by the potential barrier modulation due to Van der Waals heterojunctions in FTJ \cite{WuChen}. On top of that, the discovery of ferroelectricity in thin films of Hafnium Oxide (HfO$_2$) and zirconium doped Hafnium Oxide (H$_{0.5}$Zr$_{0.5}$O$_2$/HZO) has triggered a vast amount of researches on CMOS compatible FTJ \cite{Muelle}$^-$\cite{Trentzsch}. In the last 5-6 years, enormous studies on HZO based FTJ have demonstrated its promising future for non-volatile memories, incorporating excellent compatibility with the CMOS process\cite{Fan}$^{-}$\cite{EYang}.

Even though plenty of studies is being reported on FTJs, the domain dynamics of FTJs are still not precise yet. Therefore, a comprehensive literature review is divided into four parts to understand the domain dynamics in ferroelectric materiel.

\subsection*{Analytical Models of domain dynamics in FE}
The phenomenal paper by Bratkovsky et al. was the fundamental work that studied the impact of the insulator layer on the domain formation in a ferroelectric dielectric stack \cite{Bratkovsky}. By the minimization of total ferroelectric energy, it is shown that enhancement in the insulator layer thickness leads to decrement in domain period (i.e., number of domain increases). A recent paper by Luk'yanchuk et al. has studied the motion of domains with applied voltages \cite{Luk_yanchuk}. However, these developed models assumed that thickness of the domain wall is zero. Hence, an abrupt transition from the upward domain to the downward domain is assumed . A finite domain wall width is considered by Morozovska et al., and a variational approach is used to derive the expression of polarization profile\cite{Morozovska}. Later, Eliseev et al. proposed an analytical model which considers the finite domain wall width, and the impact of surface effect on domain formation is studied \cite{Eliseev}. However, only vertical direction gradient is considered in the polarization. 

Additionally, numerous researches also studied the domain dynamics in ferroelectric material by analyzing the domain wall velocity. The motion of domain wall by an applied electric field is classically defined by Merz law \cite{Merz}. An exponential relationship between the wall velocity and the applied electric field is established. Later, a theory of domain wall motion was developed by Miller et al.\cite{Miller} However, a discrepancy is observed between the developed approach and experiments \cite{Padilla}$^,$\cite{Tybell}. Therefore, to study the domain wall velocity at the molecular level, density functional theory (DFT) is used\cite{Grinberg}$^{-}$\cite{Liu_S}.
The impact of the dead layer on the wall velocity by the polarization screening charges is reported by numerical and analytical methods \cite{Eliseev_2}. Recently, the effect of reversed domain size over the domain wall velocity has been reported \cite{Meng}. Furthermore, a charge and heat transport-based model of domain wall motion is proposed by Yudin et al.\cite{Yudin}  
\subsection*{Numerical/Phase-Field Models of domain dynamics in ferroelectric}
Over the many years, numerical simulations or Phase-field models have played a vital role in analyzing the physics and motion of domains in the ferroelectric material.
Domain texture for a wide range of temperature is captured by numerical solutions of coupled Poisson's equation and Landau-Ginzburg equation \cite{Guerville}. The study of super-lattice (ferroelectric/paraelectric) is reported by Stephanovich et al.\cite{Stephanovich} It is demonstrated that the polarizing domains in different layers interact with each other, subsequently, calculation of critical temperature was performed. The phase-field model for micro-structural domain growth and its macroscopic effect was proposed by Zhang et al., the nucleation and motion of domain walls are studied under the impact of both electric and mechanical boundary conditions \cite{W. Zhang}. Size effect on domain pattern growth and domain switching is studied by Wang et al.\cite{Wang_J} Besides, the transition from multi-domain state to single domain state is also shown by phase-field simulations. Later, the 2-D polarization profile of multi-domains was proposed by Luk'yanchuk et al.\cite{yanchuk} The textures of soft and hard domains w.r.t. ferroelectric thickness were also studied by the authors. 
The domain texture of a thin film ferroelectric is numerically evaluated by Kontsos et al. \cite{Kontsos} Impact of strain and applied electric field on the domain evolution is shown in the paper. 

In recent years phase field models have been used for the ferroelectric-dielectric layer to study the nucleation and domain wall motion in ferroelectric with applied voltage. The phase-field model was formulated by solving the time-dependent Ginzburg–Landau (TDGL) equation with the combination of Chensky–Tarasenko (C–T) formalism\cite{Park}. Strip domain pattern is observed when the ferroelectric thickness is scaled down to a critical value. Besides, to minimize net system energy, the formation of multi-domains is observed. The Phase-field modeling approach was also applied to study nucleation and domain dynamics in a ferroelectric FET\cite{Saha_2}. The 2-D Poisson's equation coupled with Ginzburg–Landau equation is numerically solved to study the electrostatics of Metal-ferroelectric-Insulator- Metal (MFIM) and Metal-ferroelectric-Insulator-Semiconductor (MFIS) \cite{Saha_2}.
Additionally, the dynamics of soft and hard domain walls in ferroelectric heterostructures have been reported recently \cite{Saha_3}. Note that the basis of these phase field models is the minimization of total ferroelectric energy by nucleation or motion of domain walls. Competition between domain wall energy, free energy, and depolarizing energy leads to two possible scenarios: a moment of the domain wall and the nucleation of a new domain. Whether a new domain will form or the domain wall will move is determined by the interplay between various energy components, which minimizes net ferroelectric energy.
\begin{figure*}[!t]
	\centering
	\includegraphics[width=1\textwidth]{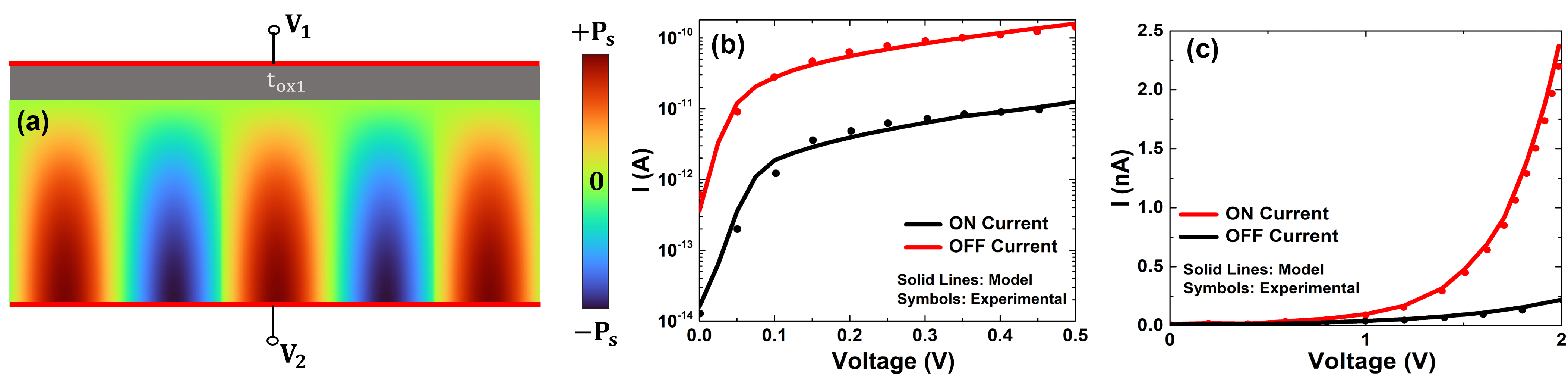}
	\caption{(a) Schematic of a Ferroelectric Tunnel Junction (FTJ) with the domain texture. (b) and (c) Validation of developed model with experimental results reported by Kuo et al\cite{Kuo_Y_s} \& Max et al\cite{B. Max_sr} respectively.
	}\label{fig:FTJ_schm}
\end{figure*}

Another phase field modeling class focuses on studying polycrystalline grains/domains in the ferroelectric material. Effect of grain boundary and its orientation on domain switching is studied by Choudhury et al.\cite{Choudhury} A review article by Chen et al. shows that the formation of domains in a polycrystalline ferroelectric takes place to reduce the total free energy and domain wall energy of the ferroelectric material\cite{Chen_L}. Liu, N et al. studies the impact of grain boundaries on the ferroelectric hysteresis loop.\cite{Liu_N} Recently, a time-dependent domain switching model of a polycrystalline ferroelectric is presented by Dasgupta et al.\cite{Dasgupta} Model was based on Poisson Voronoi Tessellation (PVT) algorithm, which is used to nucleate the grains in the ferroelectric material. Later, the PVT algorithm is also used to study the variability in the P-E loop of a 3-D multi-granular ferroelectric capacitor \cite{Pandey}. 

\subsection*{Ferroelectric Tunnel Junction Models}
An analytical model based on Tsu-Esaki's current equation reveals that the TER of an FTJ can achieve an enormous value near zero bias by the polarization switching and direct tunneling through the FE barrier\cite{Zhuravlev_1}.
Theoretical calculations of polarization reversal and restive switching in an FTJ were performed by Kohlstedt et al.\cite{Kohlstedt} Subsequently, numerical technique is used to evaluate the Tsu-Esaki current transport with depolarizing fields.
A simple 1-D model of TER is presented by Zhuravlev et al.\cite{Zhuravlev} screening effect of the metal electrode is included in the model by a simple expression of 1-D potential profile. It is shown that an increase in FE thickness leads to an enhancement in the TER ratio. 
An analytical model of tunneling transport through a FE is reported by Pantel et al \cite{Pantel}. The study considered various transport mechanisms such as Direct tunneling (DT), Fowler-Nordheim tunneling (FNT), and thermionic emission. However, the model approximated Tsu-Esaki quantum tunneling transport \cite{Davies}. A compact and explicit expression of current is derived by taking various approximations \cite{Sze}. Furthermore, a mono-domain state is assumed in the FE region. Chang et al. analyzed the TER in an FTJ, electronic transport is captured by the numerical solution of Landau formula, and polarization dynamics is included by Landau-Khalatnikov (L-K) equation \cite{Chang}. Moreover, several researchers also modeled the electron tunneling transport through a FE barrier by numerically solving the Tsu-Esaki quantum equation\cite{Tuomisto_1}$^,$\cite{Tuomisto_2}.

A numerical modeling framework for FTJ was recently reported by solving Poisson's equation with the Preisach model to include a multi-domain polarization effect \cite{Huang}. 
The mono-domain-based numerical model for metal-ferroelectric-insulator-semiconductor (MFIS) FTJ is reported by Chang et al \cite{P. Chang}$^,$\cite{P. Chang_2}. The authors analyzed both electron and hole tunneling to evaluate the ON/OFF ratio of the device. Further, a compact model for mono-domain MFIS FTJ is developed by Tung et al \cite{C. -T. Tung}. Later, Time-Dependent Landau-Ginzburg (TDLG) is solved numerically in an FTJ, and Non-Equilibrium Green Function (NEGF) is used for the quantum transport\cite{Z. Zhou}. 
\begin{figure*}[!t]
	\centering 
	\captionsetup{justification=raggedright}
	\includegraphics[width=1\textwidth]{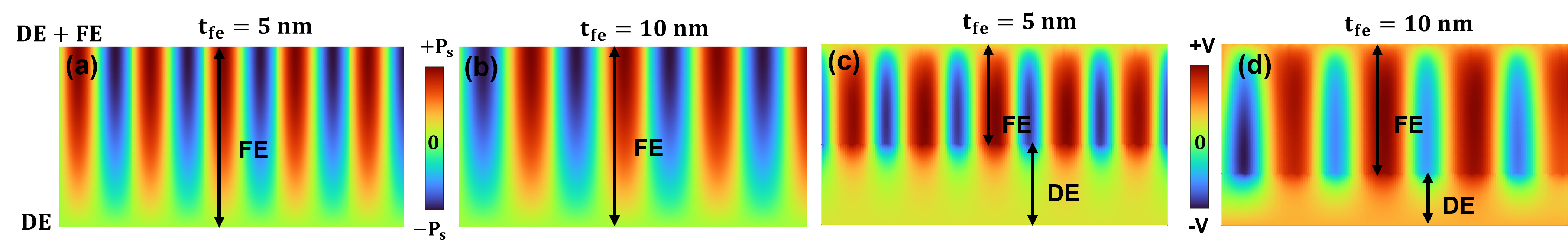}
	\caption{(a) and (b) Polarization textures for $t_{fe}$ = 5 nm and 10 nm, respectively. Scaling in $t_{fe}$ thickness leads to a denser domain pattern, and simultaneously hard to soft domain wall transitions are observed. (c) and (d) Surface plot of electrostatic potential for $t_{fe}$ = 5 nm and 10 nm, respectively. Periodicity in potential texture follows the same trend as polarization wave follows with FE thickness.
		Parameters: L = 50 nm, $t_{ox_1}$ = 4 nm and $V_2$ = $V_1$ = 0 V, $\epsilon_x$ = 22, $\epsilon_y$ = 18 and $\epsilon _{ox}$ = 10. }\label{fig:tfe_domain_zero_bias}
\end{figure*}
\begin{figure*}[!t]
	\centering 
	\captionsetup{justification=raggedright}
	\includegraphics[width=1\textwidth]{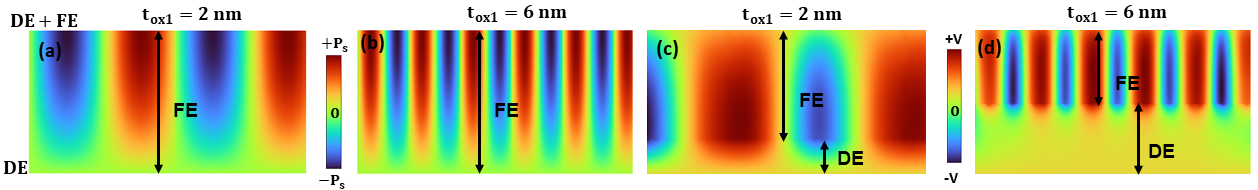}
	\caption{(a) and (b) Polarization profiles for $t_{ox_1}$ = 2 nm and 6 nm, respectively. (c) and (d) Potential distribution for $t_{ox_1}$ = 2 nm and 6 nm, respectively. Enhancement in insulator thickness corresponds to a denser domain pattern. Hence, oscillations in both polarization and potential shoot-up as $t_{ox_1}$ increases. Parameters: L = 50 nm, $t_{fe}$ = 7 nm, rest of the parameters are same as in Fig. \ref{fig:tfe_domain_zero_bias} }\label{fig:tox_domain_zero_bias}
\end{figure*}
\subsection*{Negative capacitance via domain wall motion}
R. Landauer anticipated the possibility of negative capacitance (NC) over 40 years ago \cite{Landauer}. After $\approx$ 20 years, the concept of negative permittivity (capacitance) in multi-domain FE is theoretically explained by Bratkovsky et al \cite{Bratkovsky_2}. Later, the utilization of NC to break the Boltzmann limit of sub-threshold-slope in FET is demonstrated by numerical simulations \cite{Salahuddin}. Subsequently, numerous researches have studied the NC effect in FET/capacitor either by simulations or experimental procedures\cite{Stengel}$^-$\cite{Khan_A. I_1}. 
Zukbo et al. demonstrated the NC effect via domain wall motion, the study is carried out by atomistic simulations, and comprehensive insight on the origin of NC by domain wall motion is provided \cite{Zubko}. Later, an experimental demonstration of NC by the moment of the domain wall is reported by Yadav et al\cite{Yadav}. The stability of NC states by the formation of multi-domains is discussed by Hoffmann et al \cite{Hoffmann_2}. Furthermore, in recent years, extensive focus on the origin of NC by the dynamics of domain wall motion is reported \cite{Esseni}$^-$\cite{Rollo}.
\\

After a detailed literature review of domain physics in the ferroelectric, we notice that the dynamics of domains in an FTJ are still not coherent. Therefore, in the Problem Statement section, the significance/necessity of this work is described. And the methodology to capture the necessary physics is also explained. The Results and Discussion section is focused on the impact of multi-domains on the device's electrostatics and quantum transport. The device optimization section describes the guidelines to extract the highest performance from an FTJ. Furthermore, in the supplementary material, detailed mathematics of developed model and algorithm is given, which is used in this paper. 

\section*{Problem Statement \& Methodology}
As discussed in the introduction section, in the last few years, there have been a massive number of studies are centered around FTJs. Nevertheless, only a mono-domain state in FE is considered in most research. The Preisach hysteresis model is used by Huang et al. to model multi-domains in FTJ. But, to capture the motion and nucleation of the domain wall, the total ferroelectric energy should be minimized by comprising both Landau-Ginzburg and Poisson's equation \cite{Bratkovsky}$^,$\cite{Luk_yanchuk}$^,$\cite{Morozovska}$^,$\cite{Guerville}$^,$\cite{yanchuk}$^,$\cite{Park}$^,$\cite{Saha_1}$^,$\cite{Saha_2}$^,$\cite{Saha_3}. The developed model is useful for circuit simulations but can't be used to study the domain dynamics in the FE material, which requires the coupled solutions of Poisson's equation with the Landau-Ginzburg equation. Recently, a numerical model for FTJ based on NEGF and TDLG was reported by Zhou et al \cite{Z. Zhou}. The paper shows time-dependent polarization under the pulse sequences. However, the study does not show the variations in TER or oscillations in electrostatics induced by the formation/motion of the domain wall. Besides, no information about the nucleation/motion of the domain wall due to the applied voltage is given. Therefore, the proposed study lacks to explain the oscillatory nature of electrostatics and TER/current in an FTJ.
\\

Hence, up to the author's knowledge, no study has analyzed the impact of multi-domain on electrostatics and transport characteristics of an FTJ. Fig. \ref{fig:FTJ_schm} shows a schematic of two-terminal FTJ with a top insulator (dead) layer. The fundamental problem is to capture domain dynamics in the FE region. Firstly, we solve 2-D Poisson's equation with the gradient in polarization.
\begin{align}
	\epsilon_x \frac{\partial ^2 \phi_{fe}}{\partial x^2} +\epsilon_y \frac{\partial ^2 \phi_{fe}}{\partial y^2}=-\frac{1}{\epsilon_0}\left(\frac{\partial P(x,y)}{\partial x}+\frac{\partial P(x,y)}{\partial y}\right)\; \label{poisson_fe} 
\end{align}
The 2-D solutions of the above equation are given in supplementary material. Green's function approach is generally used to obtain the electrostatic of a 2-D region bounded by Dirichlet and Neumann boundary conditions\cite{nandi}$^-$\cite{nilesh}. The next step is to incorporate the 2-D polarization wave with multi-domain. In section 1.2 of supplementary material, details about polarization modeling are presented. Firstly, we consider the equilibrium condition (zero bias condition), and the equilibrium domain period is calculated. Subsequently, the domain dynamics for non-zero applied bias is evaluated, a comprehensive discussion on methodology is given in section 1 of the supplementary material.
Note that the devolved model is explicit and analytical in nature. Therefore, we also provide the first-ever compact model for FTJ with multi-domains.

\section*{Results and Discussion}
Fig. \ref{fig:FTJ_schm}(b)\& (c) show validation of the developed FTJ model against experimental results. To check the accuracy and robustness of the developed algorithm, the model is validated against two different sets of experimental results. Fig. \ref{fig:FTJ_schm}(b) shows the ON and OFF current plots of a metal-HZO-insulator FTJ reported by Kuo et al \cite{Kuo_Y_s}, model (solid line) is thoroughly consistent with experimental data (symbols). Additionally, in Fig. \ref{fig:FTJ_schm}(c), the model is further validated with different experimental data of ON-OFF currents in an FTJ based on the HZO-Al$_2$O$_3$ FE-DE stack \cite{B. Max_sr}. Physical parameters such as metal work-function, electron affinity, FE/DE thicknesses, and permittivity are taken from the respective papers to calibrate the model with experimental results. After a rigorous calibration and validation of the developed model/algorithm, the analysis of this manuscript is categorized as follows.
\begin{figure*}[!t]
	\centering
    \captionsetup{justification=raggedright}
	\includegraphics[width=1\textwidth]{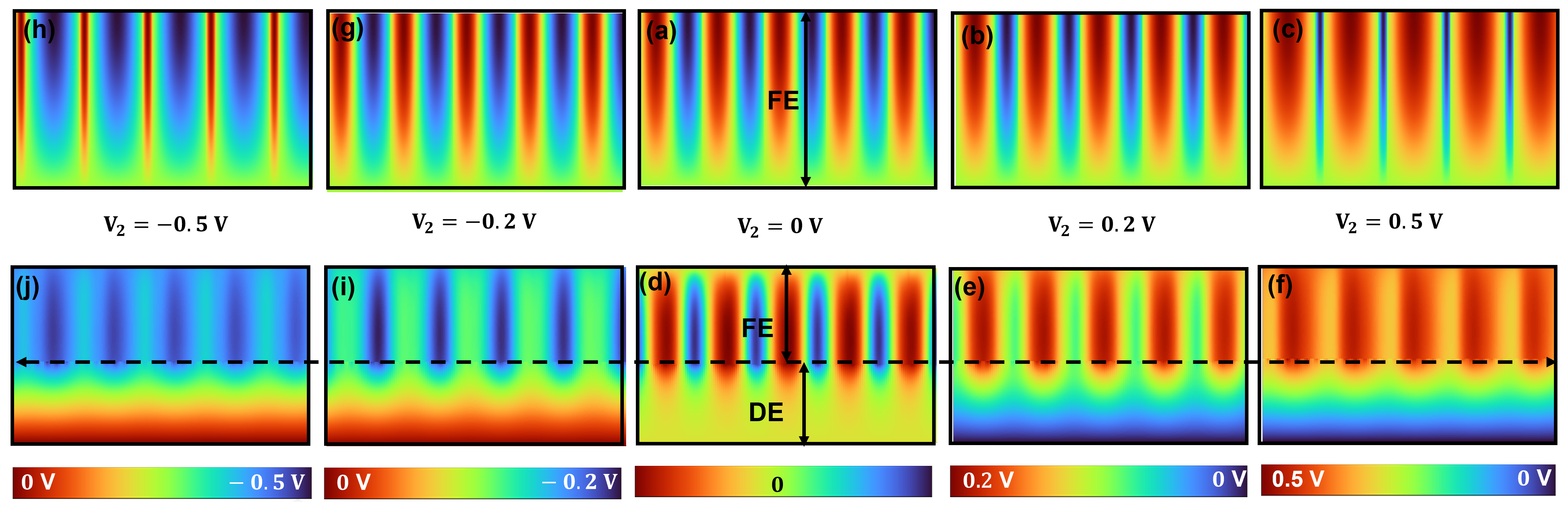}
	\caption{Potential and polarization textures for non-zero applied bias. (a)-(c) and (a)-(h) Demonstration of polarization switching with positive and negative applied bias, respectively. (d)-(f) and (d)-(j) Impact of domain switching on the electrostatic potential for positive and negative applied bias, respectively. 
	}\label{fig:non_eq_pot}
\end{figure*}
\subsection*{Electrostatics with Multi-Domain at zero bias (equilibrium condition)}
Fig. \ref{fig:tfe_domain_zero_bias} (a) and (b) show the nucleation of domain walls with variations in FE thickness ($t_{fe}$). 
An increase in $t_{fe}$ allows the domain to diffuse further along the $y-$ direction. Therefore, a domain's vertical directional gradient energy decreases ($\left ( \partial P(x,y)/{\partial y} \right ){\downarrow}$) with an enhancement in $t_{fe}$. Reduction in vertical directional gradient energy corresponds to a larger domain period because a lesser number of domains will be needed to minimize the net energy of the ferroelectric material. In other words, decrement in $\left (\frac{\partial P(x,y)}{\partial y} \right )$ reduces the gradient energy, which compensates the depolarizing energy, therefore, allows for a reduction in total system energy with a lesser number of domains. The potential profile of $t_{fe}$ = 5 nm and $t_{fe}$ = 10 nm are shown in Fig. \ref{fig:tfe_domain_zero_bias} (c) and (d) respectively, reduction in the number of domains with increased  $t_{fe}$ corresponds to lesser oscillations in potential distribution (frequency to oscillate from +V to -V).
Fig. \ref{fig:tox_domain_zero_bias} shows the domain dynamics with the alterations in dielectric layer thickness ($t_{ox1}$). In Fig. \ref{fig:tox_domain_zero_bias} (a)\& (b) domain pattern becomes denser as $t_{ox1}$ increases which is an opposite trend compared to $t_{fe}$ scaling. Reduction in domain period with increment in the $t_{ox1}$ can be explained by observing the voltage drop across the FE layer. As $t_{ox1}$ increases, voltage drop across the DE ($V_{tox_1}$) layer rises, leading to an enhancement in potential drop across the FE layer ($V_{tox_1}+V_{t_{fe}}=0$, K.V.L.). Depolarizing fields of the FE layer increases as the voltage across FE rises to compensate for increased depolarizing energy, more number of domains nucleates which minimizes the total system energy. Fig. \ref{fig:tox_domain_zero_bias} (c) and (d) show the potential profiles for $t_{ox1}$ = 2 nm and $t_{ox1}$ = 6 nm, respectively. An enhancement in $t_{ox1}$ raises the oscillations in the potential profile due to the increased number of domains.


\subsection*{Non-equilibrium Electrostatics of domains}
Fig. \ref{fig:non_eq_pot} shows the polarization and potential distribution in FE and DE regions with applied bias ($V_2$). Alteration in the magnitude of applied bias triggers the moment of the domain wall. Polarization texture with positive applied bias is shown in Fig. \ref{fig:non_eq_pot} (a)-(c). A positive enhancement in $V_2$ leads to an expansion in the upward domain period (or reduction in the downward domain period). At $V_2$ = 0.5 V, the growth in the upward domain period is substantial, and a reduction in the downward domain period is observed simultaneously. Therefore, potential distributions' oscillations (frequency to oscillate from + to - value) decline as $V_2$ increases (see Fig. \ref{fig:non_eq_pot} (d)-(f)). 

On the other hand, when applied bias is shifted along the negative direction, an expansion in the downward domain period is observed (see Fig. \ref{fig:non_eq_pot} (a), (g) \& (h)). However, oscillations in the potential profile again decrease as the $V_2$ swapped along the negative path. Therefore, all the domains will switch upward/downward at a relatively high bias, and the FE layer will attain a homogeneous potential (no oscillations) distribution.
\begin{figure}[!b]
	\centering 
	\includegraphics[width=0.5\textwidth]{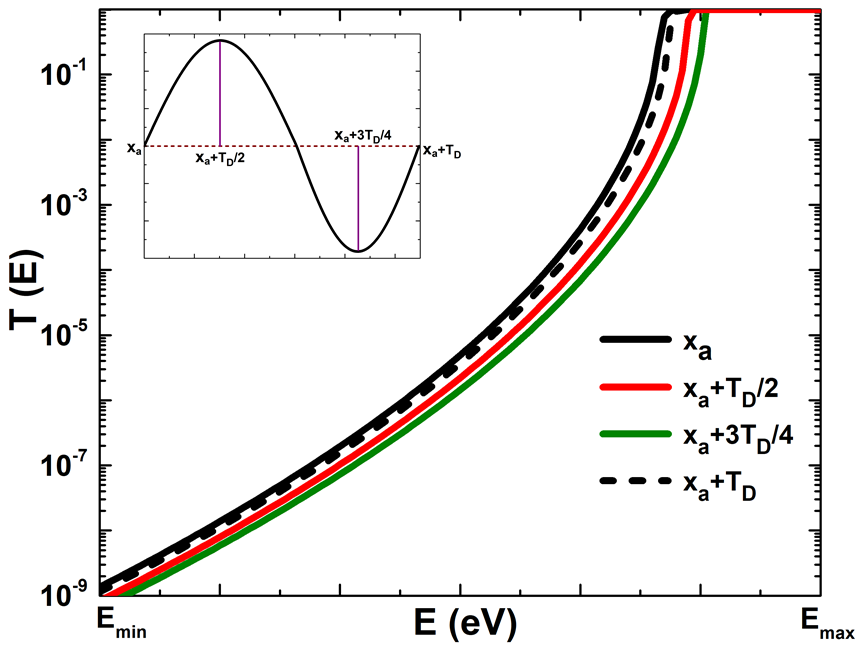}
	\captionsetup{justification= raggedright}
	\caption{Transmission probability (T(E)) of the charge carriers at various locations along the horizontal direction. Due ot the periodicity potential distributions the T(E) also exhibits an oscillatory nature. Parameters: L = 50 nm, $t_{fe}$ = 5 nm, $t_{ox_1}$ = 4 nm and $V_2$ = 0 V.}\label{fig:TE_x}
\end{figure}
\begin{figure*}[!t]
	\centering 
		\captionsetup{justification= raggedright}
	\includegraphics[width=1\textwidth]{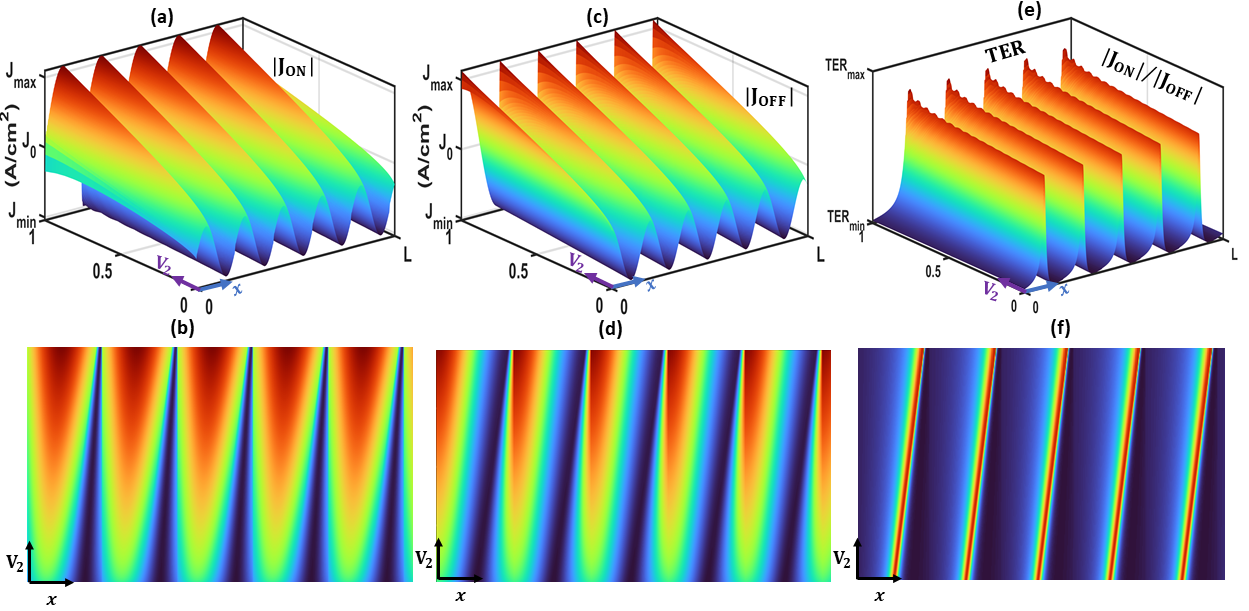}
	\caption{(a) and (b) Surface plot of the ON current density and its projection in 2-D space, respectively. (c) and (d) OFF current surface plot and its projection, respectively. Periodicity in spatial polarization charge causes the oscillations in current density. (e)-(f) The ON/OFF ratio of current density (TER) also exhibits periodic nature.
	 }\label{fig:J_ON_J_OFF_TER}
\end{figure*}
\subsection*{TER and Quantum Transports with Multi-Domain }
From Fig. \ref{fig:FTJ_schm} to Fig. \ref{fig:non_eq_pot}, it is crystalline that the formation and moment of domains in the FE region remarkably affect the device's electrostatics. Since quantum transport of electronic charge carries is originated from electrostatics. Therefore, the domain dynamics will significantly alter the device's carrier transport characteristics. This section presents a thorough analysis of quantum transport in FTJ with multi-domain.

Fig. \ref{fig:TE_x} shows electron transmission probability (T(E)) with energy at various positions along the x-direction. Position $x_a$ is the starting point of a domain, and other subsequent locations are shown in the inset of the figure. The T(E) varies enormously with different horizontal positions. Such large T(E) variations can be explained by observing Fig. \ref{fig:tfe_domain_zero_bias}(c). At the location, $x_a+T_D/2$ potential is positively maximum. 
On the other hand, at the $x_a=3+T_D/4$ position, the potential is minimum (negative), besides the difference between $+V$ to $-V$ is in the order of a few hundred mV. These huge oscillations in potential profiles lead to a spatial dependent T(E) that is extremely sensitive to the x-positions. Due to the periodicity of the domain wave, the T(E) at $x_a+T_D$ is approximately the same T(E) at  $x_a$ (see Fig. \ref{fig:tfe_domain_zero_bias}(c), potential along the x-direction is also periodic).
\begin{figure*}[!t]
	\centering 
    \captionsetup{justification= raggedright}
	\includegraphics[width=1\textwidth]{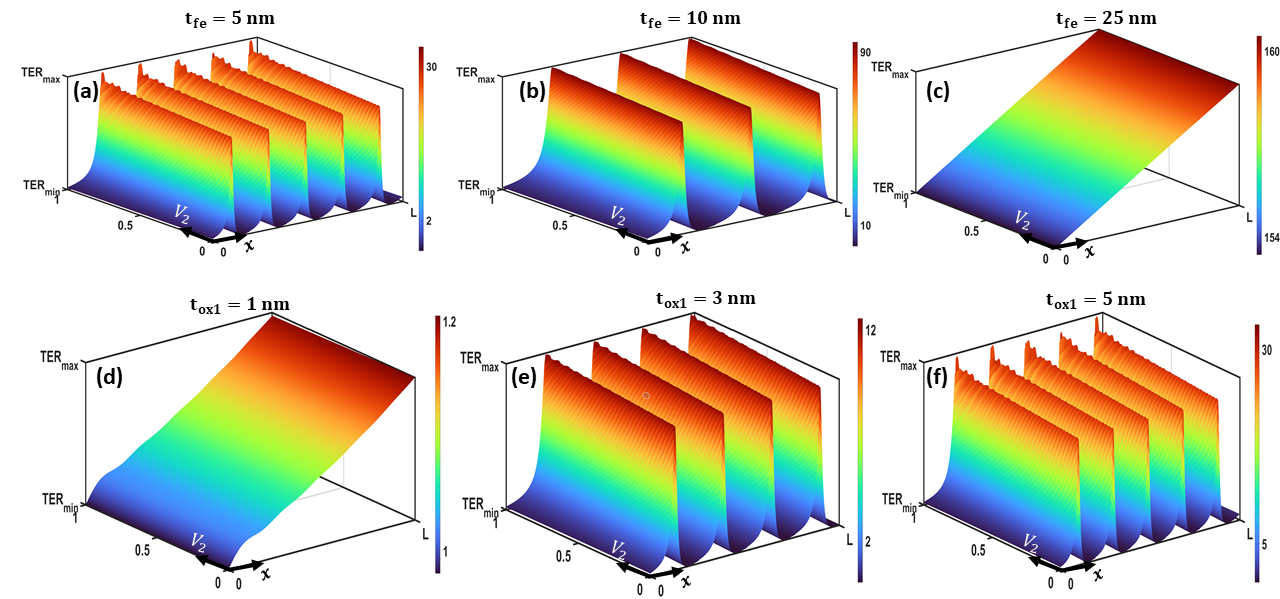}
	\caption{Optimization approaches to achieve a broader and maximum TER. (a)-(c) As ferroelectric thickness increases, the number of domains reduces, which leads to the decrement in TER fluctuations. (d)-(f) Enhancement in insulator thickness shoot up the oscillations in TER. }\label{fig:TER_tox_tfe}
\end{figure*}

Due to multi-domains, the spatial variations in T(E) will introduce local fluctuations in current density. The absolute magnitude of ON and OFF current densities are shown in Fig. \ref{fig:J_ON_J_OFF_TER} (a)-(d). Fig. \ref{fig:J_ON_J_OFF_TER} (a) shows the 3-D plot of ON current density w.r.t applied bias ($V_2$) and lateral direction (x), respectively. It is fascinating to observe that ON current density oscillates along the lateral direction. The periodicity is originated due to the oscillations in polarization charges of domains (see Fig. \ref{fig:tfe_domain_zero_bias}(a) \& Fig. \ref{fig:tfe_domain_zero_bias}(c)). The domain wall starts to move with the enhancement in applied bias at that moment the switching mechanism starts to kick in (see Fig. \ref{fig:non_eq_pot}(a)-(f)). Switching of many domains decreases the oscillations in both polarization charges as well as in potential profile (see Fig. \ref{fig:non_eq_pot}(c) \& (f)). Fig. \ref{fig:J_ON_J_OFF_TER}(b) and  Fig. \ref{fig:J_ON_J_OFF_TER}(d) show the surface plot of ON current density, as voltage rises, the domain period of the upward domain increases, which corresponds to the same amount of reduction in the downward domain period. Hence, the oscillations in current density decline by switching of domains (moment of domain wall). Depending on the device's physical parameters and value of spontaneous polarization, the ratio of $J_{max}/J_0$ can approach up to 10, which shows significant variability in the current density via dynamics of multi-domains. The OFF current density ($J_{OFF}$) is shown in Fig. \ref{fig:J_ON_J_OFF_TER}(c)-(d), a $\pi$/2 phase difference is observed between ON and OFF current densities. Therefore, if $J_{ON}$ follows the sine profile, then $J_{OFF}$ will follow cosine distribution and vice-versa.

The next step is to analyze the spatial variations in the ON/OFF ratio of the device. TER is defined as a modulation in conductivity due to the polarization reversal. Here, TER is defined as the ON/OFF current density ratio. Fig. \ref{fig:J_ON_J_OFF_TER} (e) shows 3-D variations in the TER for various combinations of $V_2$ and x-positions. As anticipated, TER also exhibits periodicity along the spatial x-direction. However, TER sharply shoots up to a peak value at some specific points. These peak points can be evaluated physically by observing minima positions of $J_{OFF}$ at which $J_{ON}/J_{OFF}$ attains its maximum value. Hence, it can be stated that if $J_{ON}$ and $J_{OFF}$ follow sine and cosine profiles, subsequently, the spatial variations in TER can be approximate as a tangent (tan(x)) function. A 2-D surface plot of TER is shown in Fig. \ref{fig:J_ON_J_OFF_TER} (f). The TER is maximum only for a narrow range of $x$ and $V_2$ combinations. But, oscillations in TER can be diminished by the interplay in domain period with FE and DE layer thickness, and a broader range of peak TER can be achieved. Hence, in the next section, we present device optimization techniques which are useful for achieving a larger TER with a broader peak.

\subsection*{Device optimization for maximum TER}
Spatial variations in TER are originated due to the periodicity in upward and downward domains (see Fig. \ref{fig:J_ON_J_OFF_TER}(e) \& (f)). And peak value of TER is obtained only for a narrow range of x-locations which is unwanted for an adequate device. A broader range of peak TER can be achieved by engineering the domain texture of FE.

Fig. \ref{fig:TER_tox_tfe} shows the optimization of TER by tuning the domain density in the FE region. Enhancement in FE thickness reduces the gradient energy (along y-direction), which relaxes the necessity of minimization of depolarization energy. Therefore, system energy can now be minimized by fewer domains. Hence, domain density decreases as FE thickness increases. At more significant $t_{fe}$, the rate of oscillations in electrostatic device decreases, therefore, the spatial variations in TER start to attenuate, and a wider range of peak TER is obtained. At $t_{fe}$ = 5 nm, the maximum to minimum TER ratio is $\approx$ 15.
On the other hand, $t_{fe}$ = 25 nm, and this ratio approaches unity due to the significantly reduced oscillations. The value of TER rises with an enhancement in the $t_{fe}$. This is due to the increased potential drop across the FE layer at a larger value of $t_{fe}$. A larger potential drop enhances the asymmetry between tunneling barriers of up and down polarization states, leading to a higher TER. Additionally, domain texture starts to move from multi-domain state to mono-domain state with increasing FE layer thickness. 
\begin{figure}[!b]
	\centering 
	\captionsetup{justification= raggedright}
	\includegraphics[width=0.5\textwidth]{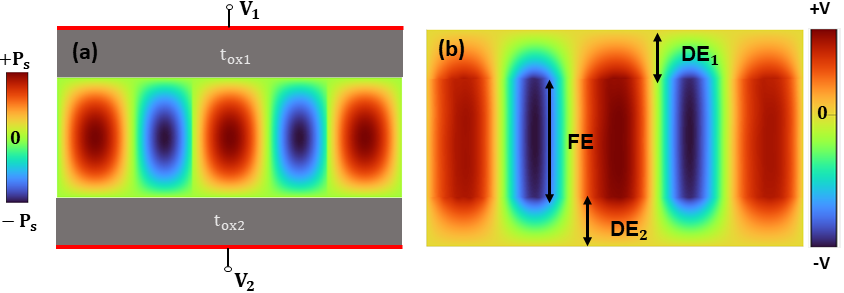}
	\caption{(a) Schematic of a multi-domain FTJ with top and bottom insulator layers. (b) Electrostatic potential distribu-  tion with multi-domain. Parameters: L = 50 nm, $t_{fe}$ = 5  nm, $t_{ox_1}$ = 3 nm, $t_{ox_2}$ = 3 nm, and $V_1$ = $V_2$ = 0 V.}\label{fig:electrostatic_tox2}
\end{figure}
\begin{figure*}[!t]
	\centering 
	\captionsetup{justification= raggedright}
	\includegraphics[width=1\textwidth]{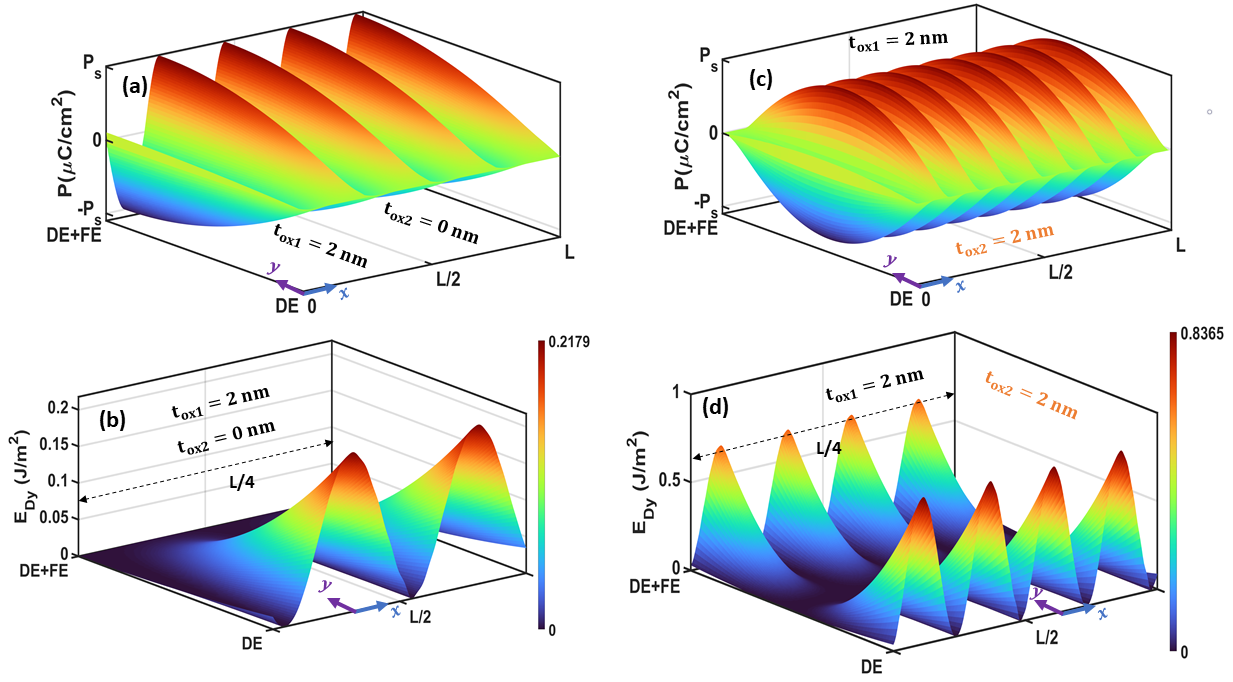}
	\caption{(a) and (b) Surface plots of polarization profile and y-directional gradient energy density for an FTJ (only with top insulator layer). (c)-(d) The FTJ with top and bottom insulator layers. (c) The incorporation of the bottom DE layer leads to a denser domain pattern. (d) Polarization wave experiences a larger gradient along the y-direction, leading to enhancement in gradient energy density.}\label{fig:E_grad_pol_tox2}
\end{figure*}
Fig. \ref{fig:TER_tox_tfe}(d)-(f) show the TER plots for various insulator thicknesses. The asymmetry in tunneling barriers for ON and OFF state polarization rises as $t_{ox_1}$ increases, leading to an enhancement in the TER value.
Nevertheless, the depolarization energy also rises with increasing $t_{ox_1}$. Hence, a denser domain pattern nucleates to minimize the depolarizing energy. Therefore, spatial variations in TER begin to shoot up as insulator thickness rises and the range of the x-positions at which TER is maximum begins to shrink. Therefore, a trade-off exists between the broader peak of TER and the maximum TER value.

\subsection*{Impact of bottom layer dielectric thickness ($t_{ox_2}$)}
Fig. \ref{fig:electrostatic_tox2} shows the polarization profile and corresponding potential surface plot of an FTJ with a bottom insulator layer ($t_{ox_2}$). Polarization is now maximum at the middle of the FE region (y = $t_{ox_1}+t_{fe}/2$) and minimum at FE/DE interfaces. Hence, FTJ with front and bottom insulator layers will have a larger polarization gradient (along the vertical direction) than FTJ with a single insulator layer. The comparison between gradient energy density and polarization profile of with and without bottom insulator layer is shown in Fig. \ref{fig:E_grad_pol_tox2}. Note that the gradient energy density of ferroelectric along y-direction ($E_{Dy}$) is higher when the bottom dead layer is incorporated (see Fig. \ref{fig:E_grad_pol_tox2}(b) and (d)). Enhancement in the $E_{Dy}$ can be attributed to the fact that now polarization profile is minimum-maximum-minimum at y = $t_{ox_1}$, y = $t_{ox_1}+t_{fe}/2$, and y = $t_{ox_1}+t_{fe}+t_{ox_2}$, respectively, which raises the gradient in the y-direction. Hence, in Fig. \ref{fig:E_grad_pol_tox2}(d) $E_{Dy}$ is higher than Fig. \ref{fig:E_grad_pol_tox2}(b).

More domains are nucleated to compensate for increased energy density due to the higher $E_{Dy}$  decreasing the depolarization energy to obtain the minima of total system energy.
But, increment in domain density decreases the domain wall width, which leads to enhancement in a gradient along the x-direction. Therefore, the enhancement in x-directional gradient energy will dominate at a certain level, and further nucleation of domains will not help minimize the total system energy. Thus, the interplay between gradient (both x and y directional), depolarization, and free energy density is crucial in determining the optimum texture of domains in the FE region.
Incorporating the bottom dead layer increases the oscillations in the polarization profile (number of domain rises). Therefore, the frequency of oscillations in potential will also go up. Hence, the variations in TER and current density will follow a similar trend as observed in electrostatics.

\section*{Conclusion}
A comprehensive study of domain dynamics in a ferroelectric tunnel junction is presented. Spatial periodicity in polarization charges induces the oscillations in device electrostatics. Since transport characteristics of devices such as tunnel electroresistance (TER), transmission probability (T(E)), and tunneling current originate from electrostatic potential, therefore, electronic transport also exhibits spatial variations. The ON and OFF current density varies $\approx$ 12 times in the transient region from the upward domain to downward domain. Due to these significant variations in current density along x-direction (horizontal-direction), the device's ON/OFF ratio (TER) also exhibits an oscillatory nature along the x-direction. The TER of the device attains its peak value only for a narrow range of x-positions and is periodically distributed throughout the ferroelectric region. To make TER maximum and uniformly along the x-direction, an optimization in domain texture is required, which can be achieved by altering the DE/FE thicknesses.
Furthermore, it is demonstrated that incorporating the bottom DE layer enhances the gradient energy (y-directional). Hence, domain texture becomes denser to compensate for increased energy density. However, the denser domain pattern will enhance the oscillatory nature of the device's electrostatics and its transport characteristics.

Additionally, a 2-D compact model of FTJ with multi-domains is formulated. The developed model can capture both phenomena domain wall motion and nucleation of a new domain. The coupled solutions of 2-D Poisson's equation (with the gradient in x-y polarization charges) and Landau-Ginzburg equation is obtained explicitly. Furthermore, the model is thoroughly validated against experimental results.



\section*{references}

\end{document}


	%
\begin{titlepage}
	\begin{center}
			\vspace*{7cm}
	
				\Huge\textbf{Dynamics of Multi-Domains in Ferroelectric Tunnel Junction}\\
				\huge\textbf{(Supplementary information)}
				
	\end{center}
\end{titlepage}
		\section{Modeling of potential landscape in FTJ with Multi-Domains}

		\subsection{Solution of 2-D Poisson's equation}
\begin{figure}[!t]
	\centering 
	\begin{minipage}{0.6\textwidth}
		\centering
		\includegraphics[width=1\textwidth]{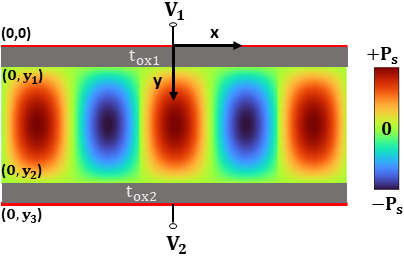}
	\end{minipage}
	\caption{Schematic of a Ferroelectric Tunnel Junction (FTJ) with multi-domains. Parameters: $t_{fe}$ = 5 nm, $t_{ox_1}$ = $t_{ox_2}$ = 3 nm, $\epsilon_{ox_1}$ = $\epsilon_{ox_2}$ 10$\epsilon_0$, $\epsilon_x$ = 22$\epsilon_0$, $\epsilon_y$ = 18$\epsilon_0$, gate metal work-function= 4.4 eV,  $V_{1}$ = $V_{2}$ = 0 V, y$_1$ = $t_{ox_1}$,  y$_2$ = $t_{ox_1}+t_{fe}$, and y$_3$ = $t_{ox_1}+t_{fe}+t_{ox_2}$.
}\label{fig:FTJ_schm}
\end{figure}

\vspace{3mm}
Fig. \ref{fig:FTJ_schm} shows the schematic of ferroelectric tunnel junction (FTJ) with multi-domains (MD) in the ferroelectric region. The 2-D Poisson's equation for the ferroelectric and dielectric layer follows as. \vspace{3mm}
	\begin{ceqn}	
			\begin{align}
			\;\;\;\;\;\;\;\;\;\;\frac{\partial ^2 \phi_{ox_1}}{\partial x^2} +\frac{\partial ^2 \phi_{ox_1}}{\partial y^2}=0\;\;\;\;\;\;\;\;\;\; 0 < y <  t_{ox_1}\label{poisson_ox1}
		\end{align}
	
		\begin{align}
\;\;\;\;\;\;\;\;\;\;\;\;\;\;\;\;\;\;\;\;\;\;	\epsilon_x \frac{\partial ^2 \phi_{fe}}{\partial x^2} +\epsilon_y \frac{\partial ^2 \phi_{fe}}{\partial y^2}=-\frac{1}{\epsilon_0}\left(\frac{\partial P(x,y)}{\partial x}+\frac{\partial P(x,y)}{\partial y}\right)\;\;\;\;\;\;\;\;\;\;  t_{ox_1} < y <  t_{ox_1}+t_{fe}\label{poisson_fe}
	\end{align}
\begin{align}
\;\;\;\;\;\;\;\;\;\;\;\;\;\;\;\;\;\;\;\;\;\;\;\;\;\;\;\;\;\;\;\;\;\;\;\;\;\;\;\;\;\;\;\;	\frac{\partial ^2 \phi_{ox_2}}{\partial x^2} +\frac{\partial ^2 \phi_{ox_2}}{\partial y^2}=0\;\;\;\;\;\;\;\;\;\;\;  t_{ox_1}+t_{fe} < y <  t_{ox_1}+t_{fe}+t_{ox_2}\label{poisson_ox2}
\end{align} \vspace{3mm}

Eq (\ref{poisson_ox1})-(\ref{poisson_ox2}) are a system 2-D partial differential equations (PDE) that possess Dirichlet and Neumann boundary conditions at the interfaces. Here, we use Green's function approach to obtain accurate and analytical solutions of potential functions for ferroelectric (FE) and dielectric (DE) regions. The first step is to derive Green's functions for various FE/DE regions. Subsequently, these functions are plugged in Green's theorem to obtain the expression of potential distribution\cite{Jackson}. Due to the uniqueness in boundary conditions (Dirichlet and Neumann), the generic solutions of a 2-D PDE system are obtained in the form of Fourier series associated with respective series indices. Green’s functions and Fourier series indices for the FE and DE regions are derived in the Appendix section.\vspace{3mm}

In 1987 Lin et al. proposed a new method to solve 2-D Poisson’s equation analytically using Green’s function technique\cite{Lin}. Later, the same group modeled short channel effects (SCE) of the SOI-MOSFET\cite{Wu}. Furthermore, in recent years, Green’s function method has been applied to solve 2-D Poisson’s and Laplace's  equations\cite{nandi}$^-$\cite{nilesh}.\vspace{3mm}

Our recent article derived an analytical model for metal-ferroelectric-insulator-semiconductor (MFIS) negative capacitance FET. The model provides coupled solutions of 2-D Poisson's equation with the Landau-Ginzburg equation\cite{nilesh}. However, a mono-domain state in the ferroelectric region is assumed. Hence, the model is not capable of capturing the effect of multi-domains.\vspace{3mm}

Therefore, in this work, the $\Delta.P$ term in (\ref{poisson_fe}) is considered to incorporate the multi-domain dynamics in the ferroelectric (FE) region. Note that both horizontal and vertical directional gradient in the polarization is incorporated. 
The model is derived for a multi-domain ferroelectric tunnel junction with top and bottom insulator (dead) layers. The 2-D potential distribution functions for the top insulator ($t_{ox_1}$) , ferroelectric ($t_{fe}$), and bottom insulator ($t_{ox_2}$) are as follows.

\vspace{5mm}
\centerline{\textbf{Top Insulator layer}}
\vspace{2mm}
\begin{align}
	\phi_{ox_1}(x,y)=-\frac{2}{L}\sum_{m}\frac{sin\left ( k_{m}x \right )D_{1}sinh( k_{m}y)}{\epsilon_{ox_1} k_{m}cos( k_{m}t_{ox_1})}+\frac{2}{L}\sum_{m}\frac{sin\left ( k_{m}x \right )cosh( k_{m}\left (t_{ox_1}-y  \right ))V_1\left ( 1+(-1)^{m+1} \right )}{ k_{m}cos( k_{m}t_{ox_1})}\nonumber \\+	\frac{2}{t_{ox_1}} \sum_{n=1}\frac{sin\left (k_{n}^{I}y  \right )\left ( A_{1}^{n}sinh\left ( k_{n}^{I}\left ( L-x \right ) \right )+A_{2}^{n}sinh\left ( k_{n}^{I}x \right ) \right ) }{ sinh( k_{n}^{I}L)}      \label{phi_ox1}
\end{align}\vspace{3mm}

\centerline{\textbf{Ferroelectric region}}
\vspace{2mm}
\begin{align}
	\phi_{fe}(x,y)=\phi_{L,R}(x,y)+\phi_{T,B}(x,y)+\phi_{MD}(x,y)+\phi_{n=0}(x)+\phi_{m=0}(x)+\phi{G_x}(x,y)  \label{phi_fe}
\end{align}\vspace{3mm}

\centerline{\textbf{Bottom Insulator layer}}
\vspace{2mm}
\begin{align}
		\phi_{ox_2}(x,y)=\frac{2}{L}\sum_{m}\frac{sin\left ( k_{m}x \right )D_{2}sinh( k_{m}\left (t_{ox_1}+t_{fe}+t_{ox_2}-y\right ))}{\epsilon_{ox_2} k_{m}cos( k_{m}t_{ox_2})}  +\frac{2}{L}\sum_{m}\frac{sin\left ( k_{m}x \right )cosh( k_{m}\left (t_{ox_1}+t_{fe}-y  \right ))V_2\left ( 1+(-1)^{m+1} \right )}{ k_{m}cos( k_{m}t_{ox_2})} \nonumber \\ +	\frac{2}{t_{ox_2}} \sum_{n=1}\frac{sin\left (k_{n}^{III}\left (t_{ox_1}+t_{fe}+t_{ox_2}-y \right )  \right )\left ( C_{1}^{n}sinh\left ( k_{n}^{III}\left ( L-x \right ) \right )+C_{2}^{n}sinh\left ( k_{n}^{III}x \right ) \right ) }{ sinh( k_{n}^{III}L)}    \label{phi_ox2}
\end{align}
All the necessary series constants and coefficients are given in the Appendix. \vspace{3mm}

The potential equation in the FE region (\ref{phi_fe}) comprises six different Fourier series terms. The methodology to derive each term and significance of the same is as follows.\vspace{3mm}

Physics and dynamics of multi-domain are primarily incorporated by the $\phi_{MD}(x,y)$ and $\phi_{n=0}(x)$ terms. The gradient in the polarization charges is responsible for the $\phi_{MD}(x,y)$ expression, which is evaluated by plugging Green's function of the ferroelectric region into the following equation. \vspace{3mm}
\begin{align}
	\phi_{MD}(x,y)=-\int_{t_{ox_1}}^{t_{ox_1}+t_{fe}}\int_{0}^{L}G_{fe}\left ( x,y;x',y' \right )\left ( \frac{\partial P(x,y)}{\epsilon_x\partial x}+\frac{\partial P(x,y)}{\epsilon_y\partial y} \right )\vspace{3mm}
\end{align}
Above identity leads to the following potential expression.\vspace{3mm}
\begin{align}
	\phi_{MD}(x,y)= \left ( \frac{2P_{s}}{t_{fe}T_D\epsilon_y } \right )\sum_{n}{cos\left (k_{n}^{II}\left ( t_{ox_1}+t_{fe}-y \right )  \right )}\sum_{j}\left ( \frac{1}{\left ( k_{n}^{II} \right )^2+k_j^2} \right )\left \{ B_j\left ( I_1^{j,n}\zeta^n k_j+I_2^{j,n}\xi^n \right )+A_j\left ( I_1^{j,n}\xi^n-I_2^{j,n}\zeta^nk_j \right ) \right \} \nonumber \\ + 2a\left ( \frac{a_0}{t_{fe}\epsilon_x } \right )\sum_{n}\frac{cos\left (k_{n}^{II}\left ( t_{ox_1}+t_{fe}-y \right )  \right )}{\left ( k_{n}^{II} \right )^2}\left \{ 1-\frac{sinh\left ( k_{n}^{II}\left ( L-x \right ) \right )+sinh\left ( k_{n}^{II}x \right )}{sinh\left ( k_{n}^{II}L \right )} \right \}\left ( \frac{1+(-1)^{n+1}}{\left ( k_{n}^{II} \right )^2} \right )	\label{phi_MD}
\end{align}\vspace{3mm}

Eq. \ref{phi_MD} contains two indices: $n$ and $j$. The polarization distribution of multi-domains is expressed in the Fourier series (see section 1.2), which is responsible for summation over $j$ in (\ref{phi_MD}). Another series over $n$ is arising due to Green's function of the FE region.\vspace{3mm}

At the limit of $n\rightarrow0$, $sinh(kx)$ can be approximate as $kx$. Hence, Green's function for the $n = 0$ term is evaluated (given in Appendix) and the corresponding potential function is derived.\vspace{3mm}

\begin{align}
	\phi_{n=0}(x)=\Lambda_1\left(\frac{P_{s}}{T_D\epsilon_xt_{fe}L }\right)\sum_{j}k_j\left ( B_jI_3^j-A_jI_4^j \right )
\end{align}\vspace{5mm}

The term $\phi_{L,R}(x,y)$ is originated due to the left (x = 0) and right (x = L) boundaries and calculated as.\vspace{3mm}
  
  \begin{align}
  	\phi_{L,R}(x,y)=	\frac{2}{t_{fe}} \sum_{n=1}\frac{cos\left (k_{n}^{II}\left ( t_{ox_1}+t_{fe}-y \right )  \right )\left ( B_1^{n}sinh\left ( k_{n}^{II}\left ( L-x \right ) \right )+B_2^{n}sinh\left ( k_{n}^{II}x \right ) \right ) }{ sinh( k_{n}^{II}L)} 
  \end{align}\vspace{3mm}
The interfaces along top-insulator/FE ($x$, $t_{ox_1}$) and FE/Bottom-insulator ($x$, $t_{ox_1}+t_{fe}$) contribute  to the $\phi_{T,B}(x,y)$ term in Eq. \ref{phi_fe}.\vspace{3mm}
\begin{align}
	\phi_{T,B}(x,y)=\frac{2}{L}\sum_{m}\frac{sin\left ( k_{m}x \right )\ }{\epsilon_{y} k_{m}sinh( k_{m}t_{fe})} \left\{D_{1}cosh\left ( k_{m}\left ( t_{ox_1}+t_{fe}-y \right ) \right )-D_{2}cosh\left ( k_{m}\left ( t_{ox_1}-y \right ) \right )\right\}
\end{align}\vspace{3mm}

$D_{1}$ and $D_{2}$ are the Fourier series coefficients and are calculated by the potential continuity conditions.
  \begin{align}
  	\phi_{ox_1}(x,t_{ox_1})=\phi_{fe}(x,t_{ox_1})
  \end{align}
 \begin{align}
	\phi_{fe}(x,t_{ox_1}+t_{fe})=\phi_{ox_2}(x,t_{ox_1}+t_{fe})
\end{align}
The expression of $D_{1}$ and $D_{2}$ is obtained as follows. 
 \begin{align}
D_1=\frac{d_{\alpha}d_5-d_{\beta}d_6}{d_5^2-d_4d_6}
\end{align}
\begin{align}
D_2=\frac{d_{\alpha}d_4-d_{\beta}d_5}{d_5^2-d_4d_6}
\end{align}\vspace{3mm} 

Interface Fourier series coefficients ($D_1$ and $D_2$) are defined in a fashion such that electric displacement field
continuity at FE/DE interface is satisfied. Thus, the developed model satisfies both 2-D potential continuity and 2-D electric displacement field continuity at the FE/DE interfaces.\vspace{3mm} 

The remaining two potential terms in (\ref{phi_fe}): $\phi_{m=0}(x)$ and $\phi_{G_x}(x,y)$ are evaluated from m = 0 and ${G_x}(x,y;x', y')$, respectively. \vspace{3mm}

\begin{align}
\phi_{m=0}(x)= \left ( \frac{P_{s}}{Lt_{fe}T_D\epsilon_y } \right )x\sum_{j}\left \{ B_j\left ( k_jI_5^{j}\Lambda_1+I_6^{j}\Lambda_2 \right ) +A_j\left ( I_5^{j}\Lambda_2-k_jI_6^{j}\Lambda_1 \right ) \right \}+ \left ( \frac{La_0}{4t_{fe}\epsilon_y } \right )x\Lambda_2
\end{align}\vspace{3mm}

\begin{align}
	&\phi {G_x}(x,y)= \nonumber \\ \nonumber 
	 &\left ( \frac{2P_{s}}{LT_D\epsilon_y } \right )\sum_{m}\frac{sin\left ( k_{m}x \right )\ }{\ k_{m}}\sum_{j}\left \{ B_j\left ( k_jI_3^{j,m}\zeta^m(y)+I_4^{j,m}\xi^m(y) \right ) +A_j\left ( I_3^{j,m}\xi^m(y)-k_jI_4^{j,m}\zeta^m(y) \right ) \right \}\\  
	&-\frac{a_0}{L\epsilon_y}\sum_{m}\frac{sin\left ( k_{m}x \right )\ }{\ k_{m}^2}\left ( 1+(-1)^{m+1} \right )\xi^m(y)
\end{align}\vspace{1mm}

All the necessary Fourier series constants are given in Appendix.
\begin{figure}[!t]
	\centering 
	\begin{minipage}{0.7\textwidth}
		\centering
		\includegraphics[width=1\textwidth]{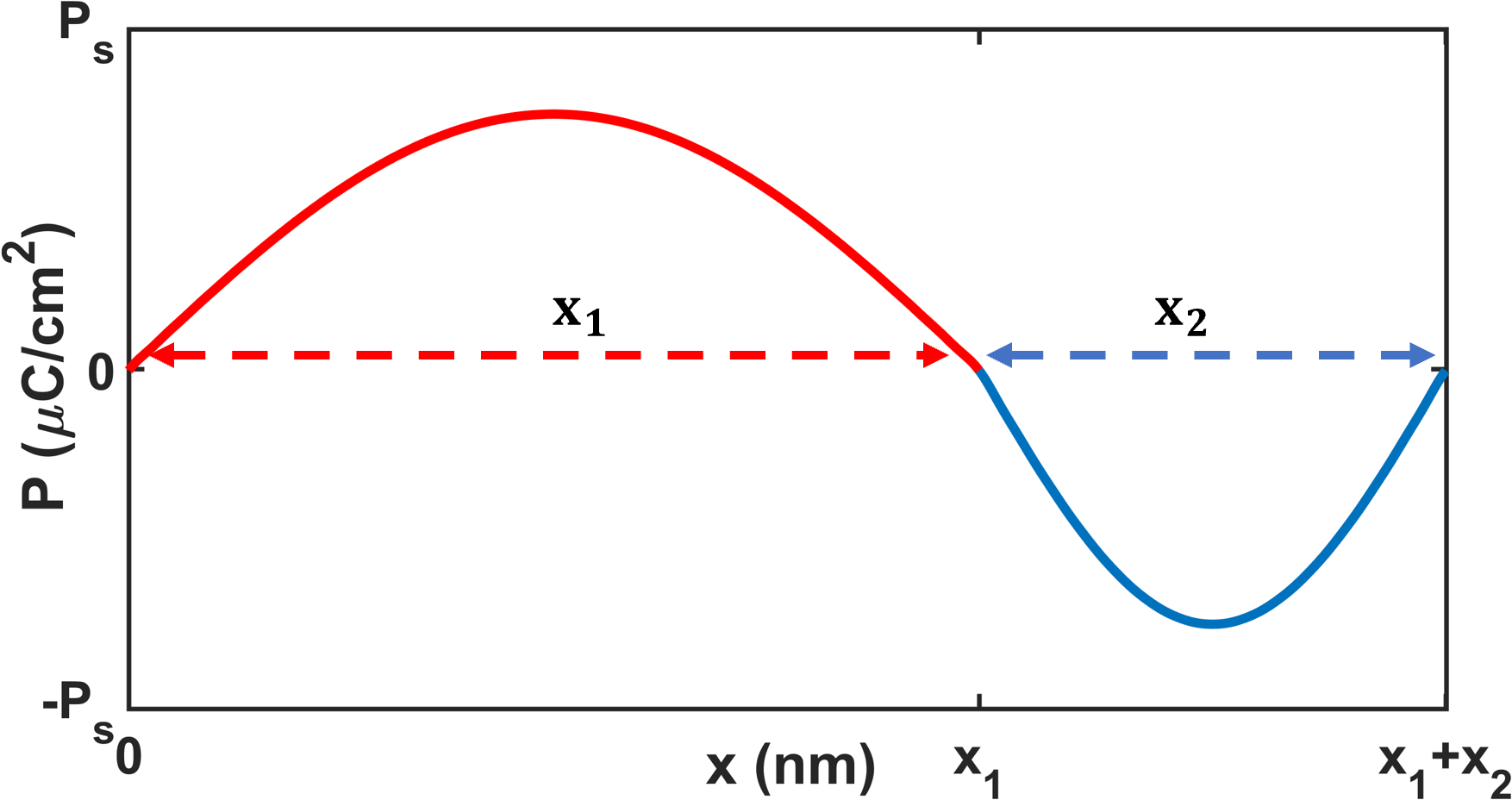}
	\end{minipage}
	\caption{Schematic of polarization wave with different $x_1$ and $x_2$ upward and downward domain periods, respectively.}\label{fig:pola_wave}
\end{figure}
\subsection{Polarization wave modeling} \vspace{3mm}

Polarization in the FE region is assumed to follow a sinusoidal wave with a period of $T_D$ and expressed in the form of the Fourier series.
	\begin{align}
	P(x,y)=f(y)\left(\frac{a_0}{2}+\frac{P_{s}}{T_D} \sum_{j}\left \{ A_jcos\left ( k_jx \right )+B_jsin\left ( k_jx \right ) \right \}\right) \label{pol_wave}
\end{align}\vspace{3mm}

where, $a_0$ is an average of Fourier wave and $A_j$, $B_j$ are Fourier series coefficients of polarization wave. The horizontal direction polarization follows a sinusoidal nature with the upward domain period $x_1$ and downward domain period $x_2$. Fig. \ref{fig:pola_wave} shows the schematic of polarization wave plotted at the mid of ferroelectric region ($y=t_{ox_1}+t_{fe}/2$). 
The vertical directional ($y$) variations in the polarization are incorporated by the function $f(y)$, which is defined by a parabolic function.
\begin{align}	
	f(y)=ay^2+by+c
\end{align}
\begin{figure}[!b]
	\centering 
	\begin{minipage}{0.6\textwidth}
		\centering
		\includegraphics[width=1\textwidth]{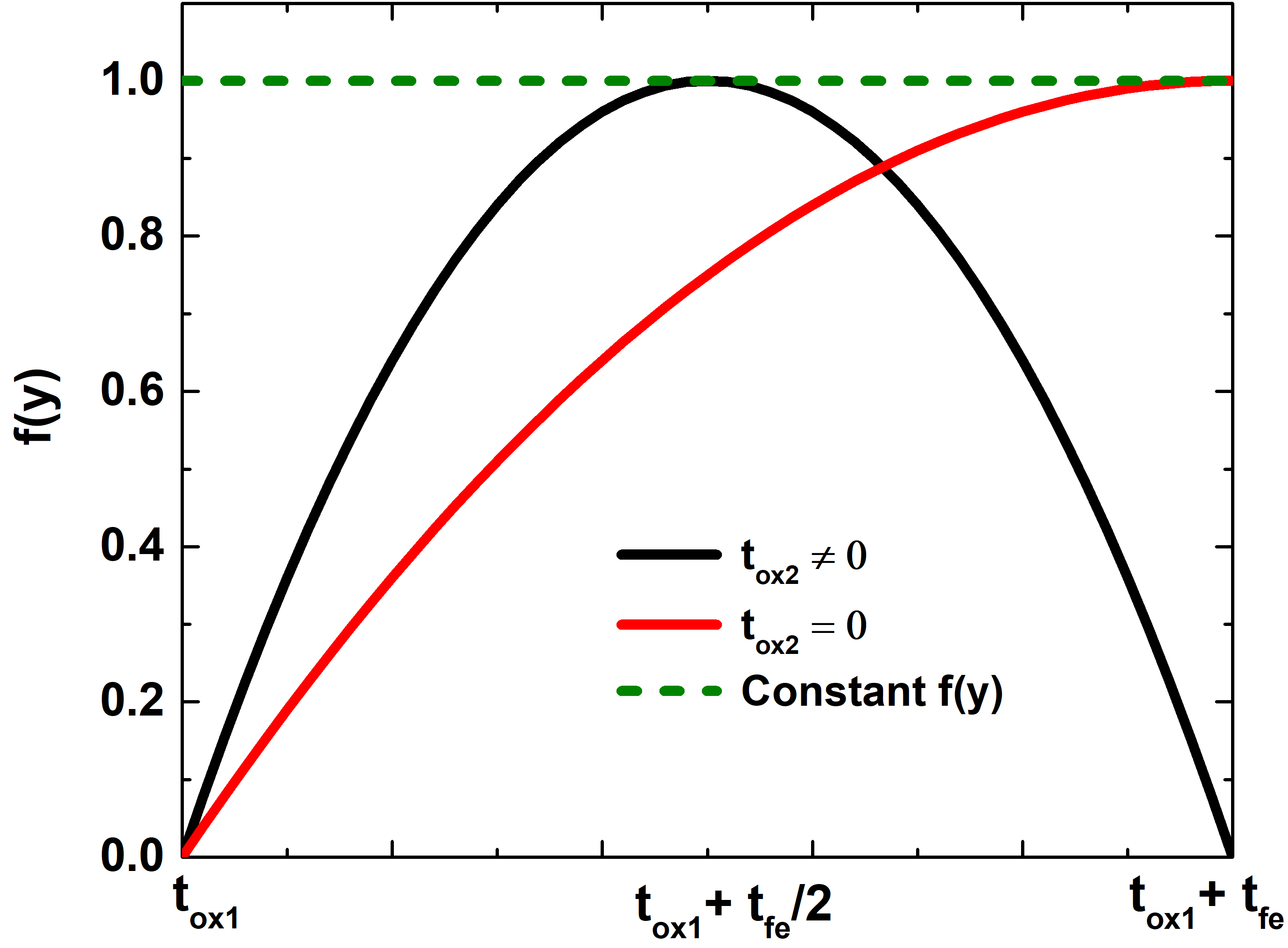}
	\end{minipage}
	\caption{Characteristics of y-directional polarization function: $f(y)$.}\label{fig:f(y)}
\end{figure}
Parabolic constants $a$, $b$ and $c$ are given in Appendix. Fig. \ref{fig:f(y)} describes the nature of the function $f(y)$ . The black curve shows an FTJ with front and back dead layers ($t_{ox_2}\neq 0$). The function takes its maximum value in the middle of the ferroelectric region ($y=t_{ox_1}+t_{fe}/2$), and it is 0 at the insulator/ferroelectric interfaces. On the other hand, the red curve plotted for $t_{ox_2}\rightarrow0$, the $f(y)$ is maximum (1) at the ferroelectric/metal interface, and it is minimum (0) at the insulator/ferroelectric interface. The green curve shows a constant $f(y)$ with no gradient along the y-direction.
\vspace{3mm}

Impact of $f(y)$ over the multi-domain polarization profile is shown in Fig. \ref{fig:pol_fy}. In Fig. \ref{fig:pol_fy}(a), polarization is maximum at the mid of ferroelectric region ($f(y)=1$) and it is zero on the ferroelectric/insulator interfaces ($f(y)=0$). Fig. \ref{fig:pol_fy}(b) shows that polarization attains its maximum value at the metal/ferroelectric interface, which is also observed in the numerical phase field simulations\cite{Saha_1}$^,$\cite{Saha_2}. In the absence of both insulator layers, it is assumed that there is no gradient in the polarization along the vertical direction (see Fig. \ref{fig:pol_fy}(c)).
\\

\begin{figure}[!t]
	\centering 
	\begin{minipage}{1\textwidth}
		\centering
		\includegraphics[width=1\textwidth]{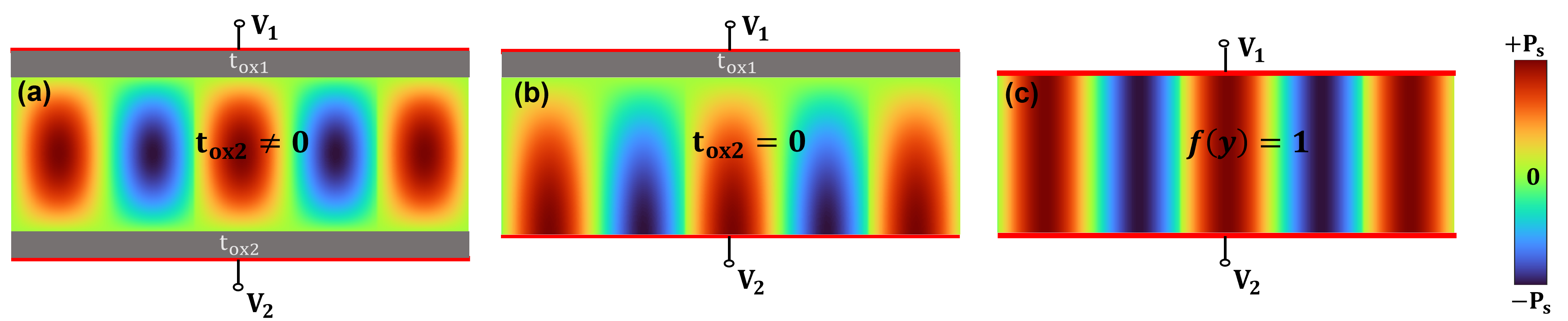}
	\end{minipage}
	\caption{Impact of $f(y)$ over the polarization profile in multi-domain FTJ. (a) Polarization pattern with top and bottom dead layers. (b) Polarization profile only with one dead layer at the top. (c) Polarization surface plot without any dead layers.}\label{fig:pol_fy}
\end{figure}
\subsection{Ferroelectric energy dynamics}\vspace{3mm}

The following equation defines the net energy density of a ferroelectric material\cite{Saha_1}$^-$\cite{Park}.\vspace{3mm}

	\begin{align}	
	 f_{_{net}} = f_{_{free}} + f_{_{dep}} + f_{_{grad}} \label{fnet}
		\end{align}\vspace{3mm}
	
	where, $f_{_{free}}$, $f_{_{dep}}$, and $f_{_{grad}}$ represent the free energy density, depolarization energy density, and gradient energy density, respectively.\vspace{3mm}
	 
The free energy density of a ferroelectric material can be expressed in the power series of polarization with associated Taylor series coefficients\cite{Lines}$^,$\cite{Rabe}.	
	\begin{align}
	 f_{_{free}} = f_0+\alpha P^2(x,y)+\beta P^4(x,y)+\gamma P^6(x,y) \label{ffree}
\end{align}\vspace{3mm}

where, $\alpha$, $\beta$, and $\gamma$ are the material-dependent parameters evaluated or taken from the experimental data used to validate/calibrate the developed model\cite{B. Max_sr}$^,$\cite{Kuo_Y_s}. The strain is not exclusive as an order parameter in (\ref{ffree}). Rather, it is included in the Landau coefficients as reported by Saha et al.\cite{Saha_1} \\

Depolarization energy of a ferroelectric material plays a crucial role in the formation of multi-domains. It is a well -established fact that to minimize depolarization energy domain formation occurs \cite{Saha_1}$^-$\cite{Rabe}. The depolarization energy density ($f_{dep}$) is obtained from the 2-D local distributions of electric fields and polarization profile in the FE region\cite{Saha_1}$^,$\cite{Saha_2}.\vspace{3mm}

		\begin{align}
		\;f_{_{dep}} = \frac{1}{2}\left[\epsilon_x \left(E^{fe}_{x}(x,y)\right)^2+\epsilon_y \left(E^{fe}_{y}(x,y)\right)^2\right]+ E^{fe}_y(x,y)P(x,y) \label{fdep}
	\end{align}\vspace{3mm}

The first term in (\ref{fdep}) shows the energy density due to spatial stray fields, and the second term signifies the contribution due to depolarizing fields in the FE region. The 2-D local electric fields are evaluated by differentiating potential function.\vspace{3mm}

\begin{align}
E^{fe}_{x}(x,y)=-\frac{\partial \phi_{fe}(x,y)}{\partial x}
\end{align}
\begin{align}
	E^{fe}_{y}(x,y)=-\frac{\partial \phi_{fe}(x,y)}{\partial y}
\end{align}\vspace{3mm}

Domain polarization exhibits gradients in both $x$ and $y$ directions\cite{Saha_1}$^-$\cite{Park}. These gradients in polarization texture cause one more energy density term defined as gradient energy density ($f_{grad}$).\vspace{3mm}

	\begin{align}
	 f_{_{grad}} = \frac{1}{2}\left[\frac{g_{11}}{2} \left(\frac{\partial P(x,y)}{\partial x}\right)^2 + \frac{g_{44}}{2}  \left(\frac{\partial P(x,y)}{\partial y}\right)^2 \right]
\end{align}\vspace{3mm}

where, $g_{11}$ and $g_{22}$ are the constants and tuned to calibrate the model with experimental results\cite{B. Max_sr}$^,$\cite{Kuo_Y_s}, note that both $x$ and $y$ directional gradients have been considered.

\section{Calculation of equilibrium domain period }\vspace{3mm}
\begin{figure}[!b]
	\centering 
	\begin{minipage}{1\textwidth}
		\centering
		\includegraphics[width=1\textwidth]{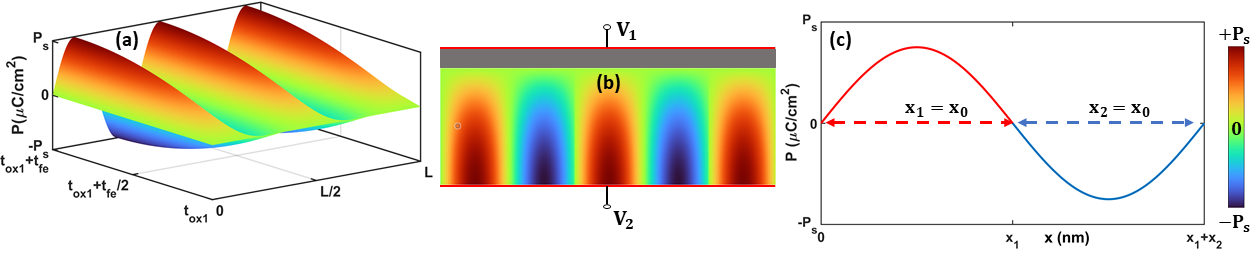}
	\end{minipage}
	\caption{Polarization distribution in the equilibrium condition. (a) 3-D distribution of polarization (b) Surface plot of $P(x,y)$ along the $x-y$ axis. Schematic of polarization wave at the equilibrium state. Parameters: $t_{ox_1}$ = 3 nm, $t_{fe}$ = 5 nm,  $t_{ox_2}$ = 0 nm and L = 30 nm, rest of the parameters are same as in Fig. \ref{fig:FTJ_schm}. }\label{fig:eq_domain}
\end{figure}

Both upward and downward domains exhibit an equal period at the zero applied voltage ($V_1$ = $V_2$ = 0), \cite{Saha_2}$^,$\cite{Bratkovsky}$^,$\cite{Luk_yanchuk}. Hence, net domain period is calculated as.

\begin{align}
	T_{D}=x_1+x_2\; (x_1=x_2=x_0) \Rightarrow T_{D}=2x_0
\end{align}\vspace{3mm}

Fig. \ref{fig:eq_domain} shows the polarization distribution in an FTJ for the equilibrium condition ($V_1$ = $V_2$ = 0). Fig. \ref{fig:eq_domain}(a) shows the 3-D polarization profile in the ferroelectric region, and Fig. \ref{fig:eq_domain}(b) shows a surface plot for the same.
Fig. \ref{fig:eq_domain}(c) shows a schematic of polarization wave with equal upward and downward domain periods ($x_0$).
The equilibrium domain period ($x_0$) is to be determined by minimizing net ferroelectric energy, which corresponds to a thermodynamically stable state\cite{Lines}$^,$\cite{Bratkovsky}$^,$\cite{Luk_yanchuk}. \vspace{3mm}

 \begin{align}
 F_{net}=\int_{t_{ox_1}}^{t_{ox_1}+t_{fe}}\int_{0}^{L}f_{_{net}}dxdy \label{Fnet}
 \end{align}\vspace{3mm}

where, $F_{net}$ is the net ferroelectric energy and $f_{_{net}}$ is calculated from (\ref{fnet}). For the given, physical and material parameters such as $t_{ox_1},\; t_{ox_2},\;t_{fe},\;L,\;\epsilon_x,\;\epsilon_y,\epsilon_{ox_1},\;\epsilon_{ox_2}$ $\alpha$, $\beta$, and $\gamma$, the only remaining unknown in (\ref{Fnet}) is $x_0$, calculated by the following equation.\vspace{3mm} 
\begin{figure}[!t]
	\centering 
	\begin{minipage}{1\textwidth}
		\centering
		\includegraphics[width=1\textwidth]{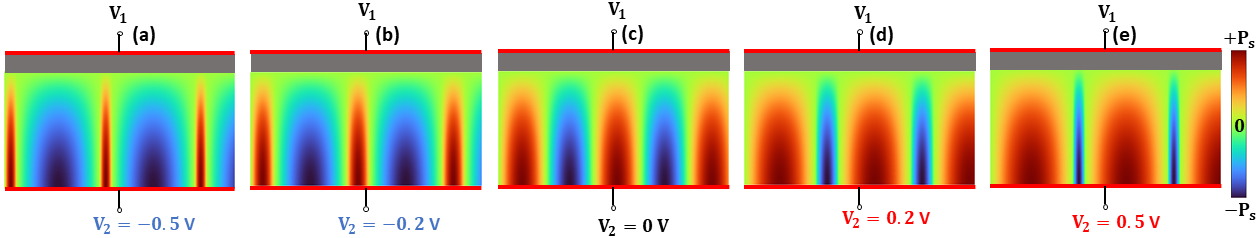}
	\end{minipage}
	\caption{Polarization distribution for non-zero applied bias. (a) -(e) Polarization distributions for $V_2$ = -0.5 V, -0.2 V, 0 V, 0.2 V and 0.5 V respectively. Parameters: $t_{ox_1}$ = 3 nm, $t_{fe}$ = 5nm,  $t_{ox_2}$ = 0 nm and L = 50 nm, rest of the parameters are same as in Fig. \ref{fig:FTJ_schm}. }\label{fig:non_eq_domain}
\end{figure}
\begin{align}
	\frac{\partial F_{net}}{\partial x_0}=0 \;(\text{solve for $x_{0}$}) \label{eq_x0}
\end{align}

Evaluated $x_0$ from (\ref{eq_x0}) is a function of physical parameters such as insulator and ferroelectric thicknesses. Discussion related to the impact of various device parameters on the equilibrium domain period is given in the manuscript's main text (see Fig. 2 and Fig. 3 of the main manuscript).
\section{Algorithm to calculate the domain period for non zero applied bias}\vspace{3mm}

Fig. \ref{fig:non_eq_domain} shows the polarization profile for a non-zero applied bias, $\left(V_2\neq 0\right)$. Due to the applied bias the equilibrium condition is no longer valid. Hence, domain periods of upward and downward domains are not identical $\left(x_1\neq x_2\right)$. For a non-zero value of $V_2$ domain wall starts to move, and depending on the polarity of voltage, an expansion or reduction in a domain period is observed. Therefore, a $\Delta x$ change in the $x_1$ leads to an equal amount of variation in the $x_2$. It is also crystalline from Fig. \ref{fig:non_eq_domain} that expansion in a domain period reduces the adjacent domain period. Hence, for $V_2\neq0$, and assuming a stationary condition (no transient) the domain period is given as.\vspace{3mm}

\begin{align}
	T_D=x_1+\Delta x+x_2 -\Delta x = x_1+x_2\Rightarrow T_D=2x_0 \label{non_eq_TD}
\end{align}\vspace{3mm}

Note that, total domain period evaluated by (\ref{non_eq_TD}) is the same as calculated by (\ref{eq_x0}). However, due to a non-zero applied bias, an asymmetry in the upward and downward domains is observed. ($x_1 \neq x_2$). The minimization of net ferroelectric energy calculates the shift in domain wall ($\Delta x$). \vspace{3mm}

\begin{align}
	\frac{\partial F_{net}}{\partial \Delta x}=0 \;(\text{solve for $\Delta x$}) \label{non_eq_x}
\end{align}\vspace{3mm}

The above equation is solved for each applied bias, and calculated $\Delta x$ is plugged in (\ref{phi_fe}) to derive the 2-D ferroelectric potential distribution incorporating multi-domains effect. The flow chart of the developed algorithm is shown in Fig. \ref{fig:flow_chart}. 
\begin{figure}[!t]
	\centering 
	\includegraphics[width=1\textwidth]{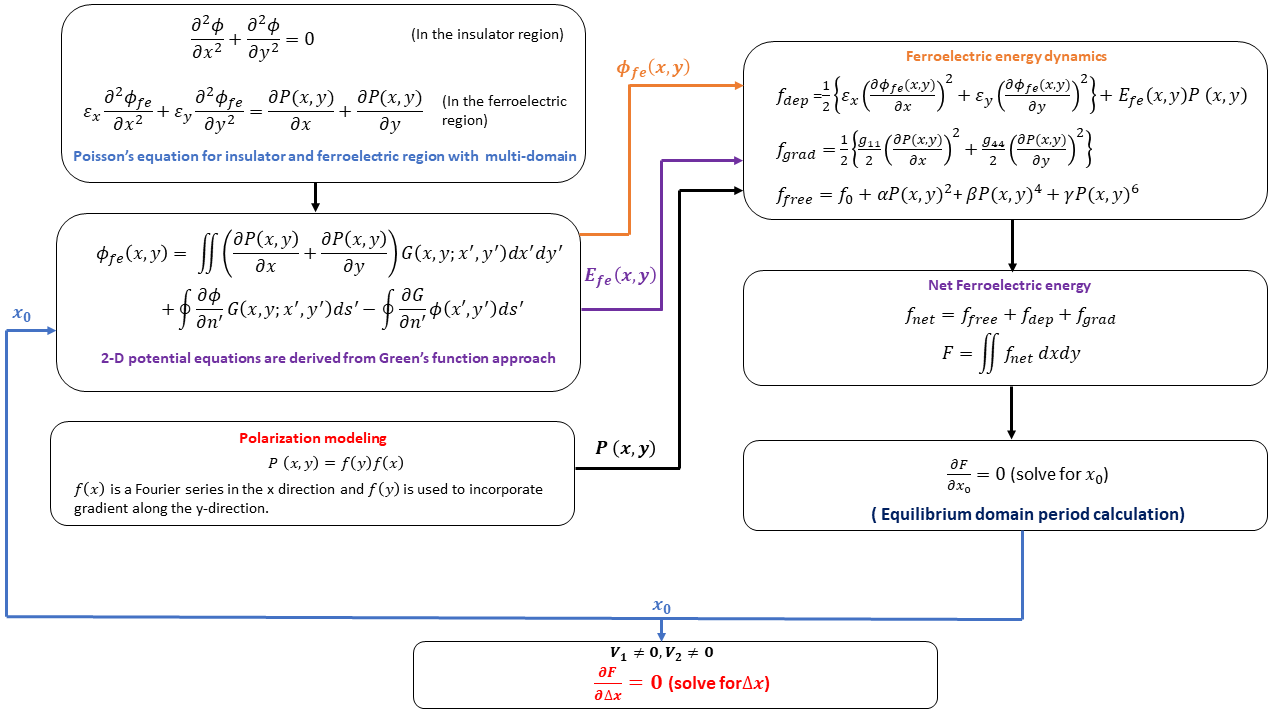}
	\caption{Flow chart of the developed model and algorithm.}\label{fig:flow_chart} \vspace{1cm}
\end{figure}
\section{Transition of Domain wall from Soft to Hard}
The phase-field simulations demonstrate that as ferroelectric thickness increases, the domain wall gradually shifts from soft wall to hard wall. These secondary effects can be incorporated into the model by molding the polarization wave. A generic expression of polarization profile is derived as follows.
\begin{align}
	P(x,y)= f(y)\left(\frac{a_0}{2}+\frac{2P_{s}}{d_wT_D} \sum_{j}\frac{A_jcos\left ( k_jx \right )+B_jsin\left ( k_jx \right )}{k_j^2}\right) \label{pol_wave_eq}
\end{align}

Fig. \ref{fig:pol_delta}(a) shows the polarization wave with a hard domain wall. A sudden change in polarization occurs, leading to an extremely high gradient along the lateral direction. Fig. \ref{fig:pol_delta}(b) shows the polarization wave texture for various $\Delta$ values. Domain wall gradually transits from soft domain wall to hard domain wall as the width of domain wall ($\Delta$) increases. Therefore, $\Delta$ can be used as a parameter to incorporate the domain wall transition. However, corresponding Fourier coefficients of polarization wave should be used in the model to capture the domain dynamics.            

\begin{figure}[!t]
	\centering 
	\includegraphics[width=1\textwidth]{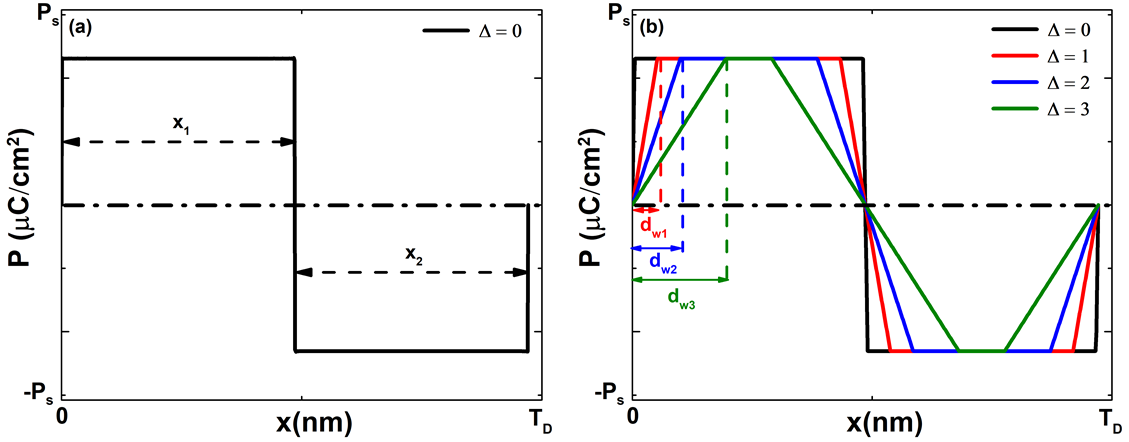}
	\caption{(a) Schematic of polarization wave with hard domain wall. (b) Gradual transition of domain wall from hard to soft domain wall.}\label{fig:pol_delta}
\end{figure}

\section{Quantum Transport Model with Multi-Domains}
The transmission probability of the charge carriers is evaluated by the Wentzel–Kramers–Brillouin (WKB) approximation.
\begin{align}
	\gamma=\frac{2\pi}{h}\sqrt{\int_{0}^{t_1}2m^{*}\left ( -q\phi (x_s,y) -E\right )}dy
\end{align}
\begin{align}
	T(E)=\exp(-2\gamma) \tag{14}
\end{align}
where, $m^{*}$ is the electron mass in DE/FE layer\cite{Vogel}, and $\phi (x_s,y)$ is the DE+FE layer potential profile at a particular $x_s$ position.
The 3-dimensional current density is calculated by a generic expression of quantum transport which is valid for all ranges of temperature\cite{Davies}.
\begin{align}
J_{T\rightarrow B}=\frac{q}{h}\left (\frac{ m^*k_BT}{\pi \hbar^2} \right )\left \{ \int_{U_L}^{\Delta \phi_B^{ox_1}} ln\left ( 1+exp\left ( E_{{fL}}-E \right ) \right )T(E)dE+\int_{\Delta \phi_B^{ox_1}}^{\infty} ln\left ( 1+exp\left ( E_{{fL}}-E \right ) \right )dE \right \} \label{current}
\end{align}

Eq. (\ref{current}) comprises both tunneling and thermionic transport of the charge carrier. The $J_{T\rightarrow B}$ denotes current from top to bottom ($t_{ox_1}$ $\rightarrow$ $t_{fe}$), $U_T$ is the reference energy point = 0, and $\mu _T$ is the electron Fermi level ($\mu _ T\approx U_T + 0.5q$)\cite{Davies}. The difference between metal work function and electron affinity of the DE layer is defined by $\Delta \phi_B^{ox_1}$. Note that the first part of (\ref{current}) is due to tunneling transport, and the second part is due to thermionic emission. Therefore, all possible current transports have been considered in the model.
In similar fashion $J_{B\rightarrow T}$ can be evaluated by replacing $U_T$, $\mu _T$, and $\Delta \phi_B^{ox_1}$ with $U_B$, $\mu _B$, and $\Delta \phi_B^{ox_2}/\Delta \phi_{FE}$ respectively, and net current density is calculated as.
\begin{align}
	J_{net}=\left(J_{T\rightarrow B}\right)-\left(J_{B\rightarrow T}\right)
\end{align}
\section{Metal Screening Effect}
The metal electrode does not entirely screen polarization charges. Hence, some residual part of these polarization charges remains unscreened in the ferroelectric layer. To incorporate such effect, the Thomas-Fermi screening approach is used\cite{Ashcroft}$^,$\cite{Pantel}. 
\begin{align}
	\Delta \phi_i=\frac{l_iQ_s}{\epsilon_0\epsilon_{M,i}  }\;\left(\text{by Pantel et al\cite{Pantel}.}\right)
\end{align}
where, $l_i$ is the screening length, $Q_s$ is screening charge density, and $\epsilon_{M}$ is the metal permittivity\cite{Pantel}. Note that, screening charge density just changes the effective applied voltage by the $\Delta \phi_i$ amount, hence, effective applied voltage is given as.
\begin{align}
	V_{e}= V_i\pm 	\Delta \phi_i
\end{align}
where, $i=1, 2$ for top and bottom metal electrodes.
\end{ceqn}
\newpage

\section{Appendix} \vspace{3mm}
Fourier series coefficients and indices which are used in the model are defined as follows.
\begin{align}
	&a=-\frac{1}{t_{fe}^2},\;b=2\left(\frac{t_{ox_1}+t_{fe}}{t_{fe}^2}\right),\;c=1-\frac{\left(t_{ox_1}+t_{fe}\right)^2}{t_{fe}^2}\\ 
	&a_0=-\frac{2P_{s}}{T_D}\left \{ \frac{cos\left ( k_2x_1 \right )\left (  cos\left ( k_2x_1 \right )- cos\left ( k_2T_D \right ) \right )+sin\left ( k_2x_1 \right )\left (  sin\left ( k_2x_1 \right )- sin\left ( k_2T_D \right ) \right )}{k_2} \right \}+\frac{4P_{s}}{T_Dk_1 }	
	\\ &k_j=\frac{2\pi}{T_D},\;k_1=\frac{\pi}{x_1},\; k_2=\frac{\pi}{x_2},\; T_D=x_1+x_2\\
	&	I_1^{j,n}=cos(k_jx)-\frac{sinh\left ( k_{n}^{II}\left ( L-x \right ) \right )}{sinh\left ( k_{n}^{II}L \right )}-\frac{cos(k_jL)sinh\left ( k_{n}^{II}x \right )}{sinh\left ( k_{n}^{II}L \right )}\\ &I_2^{j,n}=sin(k_jx)-\frac{sin(k_jL)sinh\left ( k_{n}^{II}x \right )}{sinh\left ( k_{n}^{II}L \right )} 
	\\
	&I_3^{j}=\frac{Lcos(k_jx)-xcos(k_jL)+x-L}{k_j^2},\\ 
	&I_4^{j}=\frac{Lsin(k_jx)-xsin(k_jL)}{k_j^2}
	\\
	&	I_5^{j}=\frac{Lsin(k_jL)}{k_j}+\frac{cos(k_jL)-1}{k_j^2}
	\\
	&	I_6^{j}= -\frac{Lcos(k_jL)}{k_j}+\frac{sin(k_jL)}{k_j^2}
	\\
	&	I_7^{j,m}=\left ( 1+(-1)^{m+1}cos(k_jL) \right )\frac{k_m}{k_m^2-k_j^2}
	\\
	&	I_8^{j,m}=(-1)^{m+1}sin(k_jL) \frac{k_m}{k_m^2-k_j^2}
	\\
	&I_1^{j,m,n}= \left ( 1+(-1)^{m+1}cos(k_jL) \right )\left ( \frac{k_m}{k_m^2-k_j^2}-\frac{k_m}{\left (k_{n}^{II} \right )^2+k_m^2}  \right )\\	
	&I_2^{j,m,n}= \left ( (-1)^{m+1}sin(k_jL) \right )\left ( \frac{k_m}{k_m^2-k_j^2}-\frac{k_m}{\left (k_{n}^{II} \right )^2+k_m^2}  \right )\\
	&I_3^{j,m}= \left ( 1+(-1)^{m+1}cos(k_jL) \right )\left ( \frac{k_m}{k_m^2-k_j^2}-\frac{1}{k_{m}} \right )\\
	&I_4^{j,m}= \left ( (-1)^{m+1}sin(k_jL) \right )\left ( \frac{k_m}{k_m^2-k_j^2}-\frac{1}{k_{m}} \right )\\
	&	\zeta^n= \frac{2a}{\left ( k_{n}^{II} \right )^2} \left (t_{fe}+t_{ox_1}\left ( 1+(-1)^{n+1} \right )\right)+\frac{1+(-1)^{n+1}}{\left ( k_{n}^{II} \right )^2}b \\
	&	\xi^n=2a\left ( \frac{1+(-1)^{n+1}}{\left ( k_{n}^{II} \right )^2} \right )	\\
	&	\Lambda_1=\frac{a\left \{ \left (  t_{ox_1}+t_{fe} \right )^3-t_{ox_1}^3 \right \}}{3}+\frac{b\left(\left (  t_{ox_1}+t_{fe} \right )^2-t_{ox_1}^2\right)}{2}+{c}\left({t_{fe}}\right) \\
	&	\Lambda_2=a\left \{ \left (  t_{ox_1}+t_{fe} \right )^2-t_{ox_1}^2 \right \}+bt_{fe} 
\end{align}
\begin{align}
	&	\zeta^m(y)=a\left \{ \frac{y^2}{k_m}+2\frac{t_{ox_1}cosh\left (k_{m}\left ( t_{ox_1}+t_{fe}-y \right )  \right )-\left ( t_{ox_1}+t_{fe} \right )cosh\left (k_{m}\left ( t_{ox_1}-y \right )  \right )}{k_m^2sinh\left ( k_mt_{fe} \right )} +\frac{2}{k_m^3}\right \} \\ &+b\left \{\frac{y}{k_m} +\frac{cosh\left (k_{m}\left ( t_{ox_1}+t_{fe}-y \right )- cosh\left (k_{m}\left ( t_{ox_1}-y \right )  \right ) \right )}{k_m^2sinh\left ( k_mt_{fe} \right )} \right \}+\frac{c}{k_m} \\
	&	\xi^m(y)=2a\left \{\frac{y}{k_m} +\frac{cosh\left (k_{m}\left ( t_{ox_1}+t_{fe}-y \right )- cosh\left (k_{m}\left ( t_{ox_1}-y \right )  \right ) \right )}{k_m^2sinh\left ( k_mt_{fe} \right )} \right \}+\frac{b}{k_m}
	\\ \nonumber
	&A_j=\frac{1-cos\left ( \left (k_1+k_j  \right )x_1 \right )}{k_1+k_j}+\frac{1-cos\left ( \left (k_1-k_j  \right )x_1 \right )}{k_1+k_j} -
	cos\left ( k_2x_1 \right )\frac{\ cos\left ( \left (k_2+k_j  \right )x_1 \right ) - cos\left ( \left (k_2+k_j  \right )T_D \right ) }{k_2+k_j}\\ \nonumber &-\frac{\ cos\left ( \left (k_2-k_j  \right )x_1 \right ) - cos\left ( \left (k_2-k_j  \right )T_D \right ) }{k_2-k_j}-sin\left ( k_2x_1 \right )\frac{\ sin\left ( \left (k_2+k_j  \right )x_1 \right ) - sin\left ( \left (k_2+k_j  \right )T_D \right ) }{k_2+k_j}\\ &-\frac{\ sin\left ( \left (k_2-k_j  \right )x_1 \right )  - sin\left ( \left (k_2-k_j  \right )T_D \right ) }{k_2-k_j} \\ \nonumber
	&B_j=\frac{sin\left ( \left (k_1-k_j  \right )x_1 \right )}{k_1-k_j}+\frac{sin\left ( \left (k_1+k_j  \right )x_1 \right )}{k_1+k_j} -
	cos\left ( k_2x_1 \right )\frac{\ sin\left ( \left (k_2-k_j  \right )T_D \right ) - sin\left ( \left (k_2-k_j  \right )x_1 \right ) }{k_2-k_j}\\ \nonumber &-\frac{\ sin\left ( \left (k_2+k_j  \right )x_1 \right ) - sin\left ( \left (k_2+k_j  \right )T_D \right ) }{k_2+k_j}-sin\left ( k_2x_1 \right )\frac{\ cos\left ( \left (k_2+k_j  \right )T_D \right ) - cos\left ( \left (k_2+k_j  \right )x_1 \right ) }{k_2+k_j}\\  &-\frac{\ cos\left ( \left (k_j-k_2  \right )T_D \right )  - cos\left ( \left (k_j-k_2  \right )x_1 \right ) }{k_j-k_2}
\end{align} \vspace{3mm}

A linear potential profile is assumed across the x = 0, and, x = L boundaries, which is an accurate approximation for the oxide/ferroelectric gaps \cite{pandey}$^,$\cite{nilesh}. At x = 0, the potential distribution for various region is given as. \vspace{3mm}

\begin{align}
	\phi_{ox_1}(0,y)=a_1y+a_2
\end{align}
\begin{align}
	\phi_{fe}(0,y)=b_1y+b_2
\end{align}
\begin{align}
	\phi_{ox_2}(0,y)=c_1y+c_2
\end{align}\vspace{3mm}

where, $a_1$, $b_1$ and $c_1$ are the constants calculated by the continuity condition in the electric potential and electric displacement vector as.\vspace{3mm}

\begin{align}
	\phi_{ox_1}(0,0)=V_1
\end{align}
\begin{align}
	\phi_{ox_1}(0,t_{ox_1})=\phi_{fe}(0,t_{ox_1})
\end{align}
\begin{align}
	\phi_{fe}(0,t_{ox_1}+t_{fe})=\phi_{ox_2}(0,t_{ox_1}+t_{fe})
\end{align}
\begin{align}
	\phi_{ox_2}(0,t_{ox_1}+t_{fe}+t_{ox_2})=V_2
\end{align}
\begin{align}
	-\epsilon_yb_1+P(0,t_{ox_1})=-\epsilon_{ox_1}a_1
\end{align}
\begin{align}
	\epsilon_yb_1+P(0,t_{ox_1}+t_{fe})=-\epsilon_{ox_2}c_1
\end{align}\vspace{3mm}

The Fourier series constants associated with the left (x = 0) boundary are  denoted by $A_1^{n}$, $B_1^{n}$ and $C_1^{n}$ for top insulator, ferroelectric and the bottom insulator respectively.\vspace{3mm}

\begin{align}
	A_1^n=\frac{V_1}{k_{n}^{I}}+a_1\frac{sin\left ( k_{n}^{I}t_{ox_1} \right )}{\left ( k_{n}^{I} \right )^2}
	\;;\;B_1^{n}=b_1\frac{1+(-1)^{n+1}}{\left ( k_{n}^{II} \right )^2}\;;\;
	C_1^n=c_1\left ( \frac{t_{ox_1}+t_{fe}+t_{ox_2}}{k_{n}^{III}}+\frac{(-1)^n}{\left ( k_{n}^{III} \right )^2}\right ) +\frac{c_2}{ k_{n}^{III}}
\end{align}\vspace{3mm}

Similarly the Fourier series constants at x = L, are  denoted by $A_2^{n}$, $B_2^{n}$ and $C_2^{n}$, which can be calculated in the same manner with a replacement of x = 0 by x = L.\vspace{3mm}

Fourier series coefficients for interface electric displacement vectors ($D_1$, and, $D_2$) are evaluated as follows.
\vspace{3mm}
\begin{align}
&d_{\alpha}=d_1^n+d_2^n-d_3^n-d_{pol}^{t_{ox_1}}-d_{pol}^{n=0}-d_{pol}^{m=0}-d_{pol}^{m,t_{ox_1}}
\\ 
&d_{\beta}=d_8^n+d_9^n-d_7^n-d_{pol}^{t_1}-d_{pol}^{n=0}-d_{pol}^{m=0}-d_{pol}^{m,t_1}	
\\
&d_1=\frac{V_{1}\left(1+\left(-1\right)^{m+1}\right)}{k_m cosh\left(k_m t_{ox_1}\right)}\\
&d_2=\frac{2}{t_{ox_1}}\sum_{n=1}^{\infty}\frac{sin\left(k_n^It_{ox_1}\right)k_m\left(A_1^n+A_2^n\left(-1\right)^{m+1}\right)}{k_m^2+\left(k_n^I\right)^2}\\
&d_3=\frac{2}{t_{fe}}\sum_{n=1}^{\infty}\frac{cos\left(k_n^{II}t_{fe}\right)k_m\left(B_1^n+B_2^n\left(-1\right)^{m+1}\right)}{k_m^2+\left(k_n^{II}\right)^2}\\
&d_4=\frac{1}{\epsilon_{y}k_mtanh\left(k_mt_{fe}\right)}+\frac{tanh\left(k_mt_{ox_1}\right)}{k_m\epsilon_{ox_1}} \\
&d_5=\frac{1}{\epsilon_{x}k_msinh\left(k_mt_{fe}\right)}\\
&d_6=\frac{1}{\epsilon_{y}k_mtanh\left(k_mt_{fe}\right)}+\frac{tanh\left(k_mt_{ox_2}\right)}{k_m\epsilon_{ox_2}} \\
&d_7=\frac{2}{t_{fe}}\sum_{n=1}^{\infty}\frac{\left(B_1^n+B_2^n\left(-1\right)^{m+1}\right)k_m}{k_m^2+\left(k_n^{II}\right)^2}\\
&d_8=\frac{2}{t_{ox_2}}\sum_{n=1}^{\infty}\frac{sin\left(k_n^{III}t_{ox_2}\right)k_m\left(C_1^n+C_2^n\left(-1\right)^{m+1}\right)}{k_m^2+\left(k_n^{III}\right)^2}\\
&d_9=\frac{V_{2}\left(1+\left(-1\right)^{m+1}\right)}{k_m cosh\left(k_m t_{ox_2}\right)} \\
&d_{pol}^{t_{ox_1}}=\left ( \frac{2P_{s}}{t_{fe}T_D\epsilon_y } \right )\sum_{n}{cos\left (k_{n}^{II}t_{fe}  \right )}\sum_{j}\left ( \frac{1}{\left ( k_{n}^{II} \right )^2+k_j^2} \right )\left \{ B_j\left ( I_1^{j,m,n}\zeta^n k_j+I_2^{j,m,n}\xi^n \right )+A_j\left ( I_1^{j,m,n}\xi^n-I_2^{j,m,n}\zeta^nk_j \right ) \right \} \nonumber \\ + &2a\left ( \frac{a_0}{t_{fe}\epsilon_y } \right )\sum_{n}\frac{cos\left (k_{n}^{II} t_{fe} \right )}{\left ( k_{n}^{II} \right )^2}\left \{ \left(1+\left(-1\right)^{m+1}\right)\left ( \frac{1}{k_m}-\frac{k_m}{k_m^2+\left ( k_{n}^{II} \right )^2} \right ) \right \}\left ( \frac{1+(-1)^{n+1}}{\left ( k_{n}^{II} \right )^2} \right )\\ 
&\nonumber\\ \nonumber
&\text{For $d_{pol}^{t_1}$, $cos\left (k_{n}^{II}t_{fe}  \right )$ is replaced by unity in the above equation. }
\\
&	d_{pol}^{n=0}=\Lambda_1\left(\frac{P_{s}}{T_D\epsilon_yt_{fe} }\right)\sum_{j}\frac{\left ( B_jI_3^{j,m}-A_jI_4^{j,m} \right )}{k_j} \\
&d_{pol}^{m=0}=\left ( \frac{P_{s}}{t_{fe}T_D\epsilon_y } \right )\frac{(-1)^{m+1}}{k_m}\sum_{j}\left \{ B_j\left ( k_jI_5^{j}\Lambda_1+I_6^{j}\Lambda_2 \right ) +A_j\left ( I_5^{j}\Lambda_2-k_jI_6^{j}\Lambda_1 \right ) \right \}+ \left ( \frac{\Lambda_2L^2a_0}{4t_{fe}\epsilon_y } \right )\frac{(-1)^{m+1}}{k_m}
\end{align}
\begin{align}
	&d_{pol}^{m,t_{ox_1}}= \left ( \frac{2P_{s}}{LT_D\epsilon_y } \right )\sum_{m}\frac{\left ( 1+(-1)^{m+1} \right )\ }{\ k_{m}^2}\sum_{j}\left \{ B_j\left ( k_jI_3^{j,m}\left |\zeta^m \right |_{y=t_{ox_1}}+I_4^{j,m}\left |\xi^m \right |_{y=t_{ox_1}} \right ) +A_j\left ( I_3^{j,m}\left |\xi^m \right |_{y=t_{ox_1}}-k_jI_4^{j,m}\left |\zeta^m \right |_{y=t_{ox_1}} \right ) \right \}\\ \nonumber
	&\;\;\;\;\;\;\;\;\;\;\;\;\;\;\;\;-\frac{a_0}{L\epsilon_y}\sum_{m}\frac{\left ( 1+(-1)^{m+1} \right )\ }{\ k_{m}^3}\left ( 1+(-1)^{m+1} \right )\left |\xi^m \right |_{y=t_{ox_1}}
	\\ \nonumber \\ \nonumber
&	\text{For $d_{pol}^{m,t_1}$, $\zeta^m$ and $\xi^m$ are evaluate at y = $t_1$.}
\end{align}\vspace{1mm}

The Green's functions for the DE/FE regions are defined as.\vspace{3mm}

\begin{align}
		G^{ox_1}_x(x,y;x',y')=
	\frac{2}{L}\sum_{m=1}^{\infty}\frac{sin\left(k_mx\right)sin\left(k_mx'\right)}{k_mcosh\left(k_mt_{ox_1}\right)} \begin{Bmatrix}
		sinh\left(k_my\right)cosh\left(k_m\left(t_{ox_1}-y'\right)\right)\;\;\;\;\; y<y'& \\ 
		sinh\left(k_my'\right)cosh\left(k_m\left(t_{ox_1}-y\right)\right)\;\;\;\;\; y>y'& 
	\end{Bmatrix}	
\end{align}
\begin{align}
	G^{ox_1}_y(x,y;x',y')=
	\frac{2}{t_{ox_1}}\sum_{n=1}^{\infty}\frac{sin\left(k_n^Iy\right)sin\left(k_n^Iy'\right)}{k_n^Isinh\left(k_n^IL\right)}
	 \begin{Bmatrix}
		sinh\left(k_n^Ix\right)sinh\left(k_n^I\left(L-x'\right)\right)\;\;\;\;\; x<x'  & \\ 
		sinh\left(k_n^Ix'\right)sinh\left(k_n^I\left(L-x\right)\right)\;\;\;\;\; x>x' & 
	\end{Bmatrix}	
\end{align}
\begin{align}
	G^{fe}_x(x,y;x',y')=
	\frac{2}{L}\sum_{m=1}^{\infty}\frac{sin\left(k_mx\right)sin\left(k_mx'\right)}{k_msinh\left(k_mt_{fe}\right)} \begin{Bmatrix}
		cosh\left(k_m\left(t_{ox_1}-y\right)\right)cosh\left(k_m\left(t_1-y'\right)\right)\;\;\;\;\; y<y'  & \\ 
		cosh\left(k_m\left(t_{ox_1}-y'\right)\right)cosh\left(k_m\left(t_1-y\right)\right)\;\;\;\;\;\; y>y' & 
	\end{Bmatrix}	
\end{align}
\begin{align}
	G^{fe}_y(x,y;x',y')=
	\frac{2}{t_{fe}}\sum_{m=1}^{\infty}\frac{cos\left(k_n^{II}\left(t_1-y\right)\right)cos\left(k_n^{II}\left(t_1-y'\right)\right)}{k_n^{II}sinh\left(k_n^{II}L\right)} \begin{Bmatrix}
		sinh\left(k_n^{II}x\right)sinh\left(k_n^{II}\left(L-x'\right)\right)\;\;\;\;\; x<x'  & \\ 
		sinh\left(k_n^{II}x'\right)sinh\left(k_n^{II}\left(L-x\right)\right)\;\;\;\;\; x>x' & 
	\end{Bmatrix}	
\end{align}
\begin{align}
	G^{ox_2}_x(x,y;x',y')=
	\frac{2}{L}\sum_{m=1}^{\infty}\frac{sin\left(k_mx\right)sin\left(k_mx'\right)}{k_mcosh\left(k_mt_{ox_2}\right)} \begin{Bmatrix}
		cosh\left(k_m\left(t_1-y\right)\right)sinh\left(k_m\left(t_2-y'\right)\right)\;\;\;\;\; y<y'& \\ 
		cosh\left(k_m\left(t_1-y'\right)\right)sinh\left(k_m\left(t_2-y\right)\right)\;\;\;\;\; y>y'& 
	\end{Bmatrix}	
\end{align}
\begin{align}
	G^{ox_2}_y(x,y;x',y')=
	\frac{2}{t_{ox_2}}\sum_{n=1}^{\infty}\frac{sin\left(k_n^{III}\left(t_2-y\right)\right)sin\left(k_n^{III}\left(t_2-y'\right)\right)}{k_n^{III}sinh\left(k_n^{III}L\right)}
	\begin{Bmatrix}
		sinh\left(k_n^{III}x\right)sinh\left(k_n^{III}\left(L-x'\right)\right)\;\;\;\;\; x<x'  & \\ 
		sinh\left(k_n^{III}x'\right)sinh\left(k_n^{III}\left(L-x\right)\right)\;\;\;\;\; x>x' & 
	\end{Bmatrix}	
\end{align}\vspace{3mm}

where, $t_1=t_{ox_1}+t_{fe}$, $t_2=t_{1}+t_{ox_2}$, $k_m=m\pi/L$, $k_n^{I}=(2n-1)\pi/(2t_{ox_1})$, $k_n^{II}=n\pi/t_{fe}$, and, $k_n^{III}=(2n-1)\pi/(2t_{ox_2})$.